# ThRIve: Thermally Robust CNN Inference via Low-Rank Adaptation in Heterogeneous PIM Architectures

Vibhanshu Sharma

Washington State University, vibhanshu.sharma@wsu.edu

Pratyush Dhingra

Washington State University, pratyush.dhingra@wsu.edu

Janardhan Rao Doppa

Washington State University, jana.doppa@wsu.edu

Partha Pratim Pande

Washington State University, pande@wsu.edu

Processing-In-Memory (PIM) has emerged as a promising technology for accelerating machine learning (ML) workloads. Specifically, non-volatile memory-based PIM architectures have enabled effective ML acceleration due to their ability to perform energy-efficient matrix-vector multiplication operations. However, these devices suffer from non-idealities such as thermal noise. This noise alters the stored values in the memory cells which correspond to actual model weights, compromising the inference accuracy. In this work, we introduce ThRIve, a noise-aware training methodology that leverages low-rank adaptation to enable thermally robust inference on heterogeneous PIM architectures. ThRIve selectively stores these low-rank noise-aware parameters on a hardware that is less susceptible to thermal noise, enabling robustness against temperature-induced noise variations. ThRIve mitigates the effects of thermal-noise and prevent the drop in inference accuracy across the entire operating temperature range. Experimental results demonstrate that ThRIve-enabled architectures maintain consistent inference accuracy, with the mean accuracy staying within 2% of the ideal (i.e., noise-free) accuracy, and the variation in accuracy across the entire operating temperature range remaining within 2% of the mean. The proposed methodology achieves accuracy and robustness comparable to thermally-resilient Static Random-Access Memory (SRAM)-based PIM systems, while delivering up to 5.4× reduction in energy-delay product (EDP) during CNN model inferencing.

CCS CONCEPTS • Hardware → Emerging technologies • Hardware →Robustness → Hardware reliability • Computing methodologies → Machine learning

**Additional Keywords and Phrases:** PIM, CNNs, Thermal noise-aware training, machine learning, low-rank adaptation

**ACM Reference Format:**

Vibhanshu Sharma, Pratyush Dhingra, Janardhan Rao Doppa, and Partha Pratim Pande. 2025. ThRIve: Thermally Robust CNN Inference via Low-Rank Adaptation in Heterogeneous PIM Architectures: ACM Conference Proceedings Manuscript. ACM Trans. Des. Autom. Electron. Syst. 31, 2, Article 23 (December 2025), 25 pages. https://doi.org/10.1145/3774328

This work was supported by the US National Science Foundation grant CSR-2308530 and by the Army Research Office grant ARO-W911NF-24-1-0240.

Authors' Contact Information: Vibhanshu Sharma (corresponding author), e-mail: vibhanshu.sharma@wsu.edu; Pratyush Dhingra, e-mail: pratyush.dhingra@wsu.edu; Janardhan Doppa, e-mail: jana.doppa@wsu.edu; Partha Pratim Pande, e-mail: pande@wsu.edu; The authors are with the EECS department, Washington State University, Pullman, Washington, United States.

## 1 INTRODUCTION

As we transition towards an era of AI-powered edge applications, it has become crucial to explore hardware architectures that are more energy-efficient than conventional GPUs, while still meeting the performance requirements necessary to execute the data-intensive Machine Learning (ML) workloads [1], [2]. Processing-in-Memory (PIM), which enables energy-efficient Matrix-Vector Multiplication (MVM), has emerged as a promising technology for accelerating ML workloads [3], [4], [5]. PIM systems perform computations directly within the memory, reducing unnecessary data movement and helping to overcome the memory-wall challenge [3], [4], [6]. PIM implementations employ both CMOS-based volatile memory such as Static Random-Access Memory (SRAM) and Non-Volatile Memory (NVM) devices such as Resistive Random-Access Memory (ReRAM), Phase Change Memory (PCM), and Ferroelectric Field-Effect Transistors (FeFET) etc. [7], [8], [9], [10], [11]. NVM devices offer significantly higher storage density compared to SRAM-based volatile memory devices [12], [13]. This allows considerably large amount of memory to be integrated on-chip, making them a popular choice for ML deployment at the edge [13], [14].

Though NVM-based PIM architectures are superior compared to conventional Von Neuman architectures such as GPUs in terms of energy-efficiency and performance, these devices suffer from non-idealities [15], [16]. Various non-idealities, such as IR-drop, random telegraph noise, shot noise, programming variations, stuck-at-faults, quantization loss, and thermal noise can negatively affect ML model's accuracy [12], [13], [14], [17], [18]. Specifically, these non-idealities alter the stored values in the memory cells that correspond to actual weights, compromising the accuracy [14], [19]. To address this challenge, several methods have been proposed so far that aims to mitigate the negative effects of these non-idealities on predictive accuracy [14], [15], [16], [19], [20].

Thermal noise, an inherent challenge in NVM-based PIM devices, poses a significant obstacle [21], [22]. These architectures suffer from reduced inference accuracy due to their high temperature sensitivity [23], [24]. To address this challenge, noise-aware training (NAT) techniques have been developed. A state-of-the-art NAT technique utilizes student-teacher framework to enhance robustness against thermal noise [25]. This approach combines two strategies: Progressive Knowledge Distillation (PKD) and Batch Norm Adaptation (BNA). In the PKD method, a neural network (student) is incrementally trained with higher temperature noise levels. At each level, the previously trained student serves as the new teacher model, distilling its knowledge of the noise into the next student model. In the BNA strategy, temperature-specific normalization parameters of student are trained while the network weights remain frozen [25]. However, these NAT methods fail to provide robustness across the entire temperature range in NVM-based PIM systems. There are two primary limitations with such methods: (i) these methods perform offline NAT on GPU/CPU at specific temperature values. As an example, two temperature values (25°C,35°C) are chosen for full parameter NAT using PKD, while three (55°C,85°C,120°C) are chosen for BNA. (ii) loading NAT-trained parameters onto an NVM crossbar array exposes the weights to further noise that needs to be addressed [25]. Together, these limitations hinder the effectiveness of NAT to restore inference accuracy across the entire temperature range. The noise-aware trained model lacks generalizability, recovering poorly at unseen temperatures. It provides robustness only at the temperature values used during training. Moreover, frequent retraining is impractical for deployed ML models due to the limited write endurance of NVM devices [26].

To overcome the limitations of existing NAT techniques for thermally-robust inference using NVM-based PIM devices, it is necessary to store the noise-aware parameters obtained after NAT on a hardware that is less susceptible to thermal noise. However, in current NAT techniques, this would mean storing all model parameters in a thermal noise-resilient environment. Such storage demands would lead to significant hardware overhead solely to ensure thermal robustness. To address this issue, we propose ThRIve, a novel NAT framework designed to maintain consistent inference accuracy across the entire operating temperature range and to enable thermally-robust inference in heterogeneous NVM-CMOS PIM architectures. ThRIve reduces storage overhead by using Low-Rank Adaptation (LoRA), which introduces small, trainable low-rank matrices to modulate the pretrained model's weights and has been successfully applied for fine-tuning large language models [27]. During the NAT stage in our ThRIve framework, the low-rank matrices adapt to changing noise distributions through backpropagation, while the original model parameters remain frozen. At deployment, only these LoRA parameters are stored on a hardware that is less susceptible to thermal noise. This approach significantly reduces hardware overhead, as the LoRA parameters represent only a small portion of the full model. CMOS-based hardware offers excellent noise resiliency, and several prior studies have demonstrated its effectiveness when combined with NVM devices [12], [13], [14], [19]. Accordingly, we leverage existing NVM-CMOS heterogeneous PIM architectures that integrate NVM cell technologies such as ReRAM, PCM, and FeFET with SRAM, a CMOS-based memory, to support ThRIve-enabled inference.

Furthermore, to avoid the inefficiency of training across all thermal noise levels, ThRIve incorporates a Bayesian Optimization-driven NAT framework. This method adaptively searches the given input space, which is defined by the temperature-dependent noise distribution to select a representative subset of noise levels. The LoRA matrices are then trained on this subset, enabling efficient yet thermally-robust inference across a wide operating temperature range. This approach enables reliable, and thermally-resilient inference for ML models on NVM-CMOS based heterogeneous PIM systems. The key contributions of this paper are as follows:

1. We propose to incorporate LoRA within heterogeneous NVM-CMOS architectures to enhance thermal robustness of the PIM systems executing ML inference workloads. By using LoRA, we significantly reduce the number of additional parameters needed to be stored on SRAM to provide thermal robustness, while keeping the original model parameters on NVM. We demonstrate the applicability of the proposed method using well-known existing heterogeneous PIM architectures to improve their resilience to thermal noise.
2. We develop an efficient Bayesian optimization-driven NAT methodology where LoRA's trainable low-rank matrices adapt to varying noise distributions via backpropagation and generalizes to unseen temperatures. This approach ensures robustness without altering the base model parameters, removes the need for frequent on-device retraining, and reduces write operations to NVM cells, thereby resulting in an improved lifespan.
3. Our experimental results demonstrate that ThRIve outperforms existing noise-aware training methods in terms of inference accuracy over the entire operating temperature range (generalized prediction accuracy). ThRIve enabled architectures maintain consistent inference accuracy, with the mean within 2% of the ideal accuracy (achieved in a noise-free environment), and the variation across the entire temperature range remaining within 2% of the mean value. It helps the heterogeneous architecture to achieve accuracy and thermal robustness comparable to thermal noise resilient SRAM-based PIM systems, while delivering up to ~5.4× reduction in energy-delay product (EDP) during CNN model inferencing.

The rest of this paper is organized as follows. Section-2 discusses the related prior work, Section-3 presents the problem formulation. Section-4 elaborates about the overall framework and proposed methodology. Section-5 presents the experimental analysis and finally Section-6 concludes the paper by summarizing the key features.

## 2 RELATED WORK

In this section, we first discuss about how PIM devices are used for accelerating ML and existing PIM architectures. Next, we discuss the existing noise mitigation techniques.

### 2.1 PIM Architectures and ML Acceleration on PIM

PIM architectures perform computations directly in memory arrays. This reduces data movement between storage and processing units. NVM crossbars in PIM systems leverage Ohm's and Kirchhoff's current laws to perform MVM operations. Specifically, an $N \times N$ crossbar can perform $N^2$ multiply-accumulate operations simultaneously in $O(1)$ time, making these systems extremely energy efficient and enabling high performance [28], [29]. The computational kernels of various types of deep neural networks (DNNs) can be broken down into MVM operations, that can be efficiently implemented on PIM architectures [30], [31], [32] . By leveraging parallelism within memory arrays, PIM architectures can offer significant speedup and superior energy-efficiency over GPUs while executing ML workloads [30], [33]. PIM-based architectures are implemented using a wide range of memory technologies, each with unique advantages and challenges [3], [29]. NVM technologies such as ReRAM offer high storage density, small cell area, and low leakage energy [13]. PCM devices provide similar cell area benefits as ReRAM with better retention characteristics [12], [13]. FeFET devices have a slightly larger cell area compared to ReRAM and PCM but provide low write energy, high read/write speed, and improved thermal resilience [13]. Despite this, FeFET cells have limited write endurance and poor retention, necessitating frequent reprogramming and increasing inference energy [11], [12]. However, all ReRAM, PCM and FeFET cells are susceptible to thermal noise, thereby leading to a degradation in the predictive accuracy of DNNs [24], [34], [35]. Volatile memory options such as SRAM and DRAM come with their own strengths and weaknesses. For instance, SRAM is robust against thermal noise, and it has low write latency [13], [19]. However, it has a large cell area and low storage density (i.e., can only store 1-bit per-cell) [12], [13]. DRAM, on the other hand, offers a smaller cell size but comes with high leakage power, frequent refresh requirements, and limited in-situ computing capability [13], [36]. There are several existing PIM architectures that rely solely on NVM technologies for DNN training/inference [4], [10], [37]. In addition, heterogeneous PIM architectures have also been developed to address the limitations of homogeneous PIM architectures [12], [13], [14], [19].

### 2.2 Thermal noise mitigation techniques for PIM devices

NVM devices are susceptible to thermal variations, which can cause changes in stored weights during DNN model training/inference. This poses a significant challenge for maintaining the reliability and usefulness of NVM-based PIM devices [21], [23], [25]. Prior works have tried to address the effect of thermal noise by employing various noise mitigation techniques. These techniques can be broadly categorized into hardware, software, and hybrid approaches. Hardware approaches involve additional circuitry to counteract non-idealities, though this often introduces area and performance overhead to the base architecture [21], [38]. A ReRAM/SRAM PIM architecture has been proposed, which combines the outputs of a noisy ReRAM macro and a precise SRAM macro using a programmable shifter to enable bit-level compensation. This approach mitigates non-idealities such as device-to-device variations, limited resistance levels, etc. [14]. However, despite introducing significant SRAM area overhead, this method is unfit for thermal noise mitigation as temperature is a continuous variable, and the programmable shifter supports only discrete compensation levels. Another work identifies important input channels using a Hessian-based analysis [39]. These channels, along with their corresponding input data, are then moved to a noise-resilient SRAM-based PIM accelerator, thereby reducing variation-induced accuracy degradation [39]. Additionally, a method based on temperature-aware row assignment has been developed to remap DNN weights away from high-temperature rows in the crossbar [38]. However, this weight remapping leads to increased ReRAM updates, impacting its lifetime. Furthermore, methods such as weight reordering, row adjustment and error compensation have also been explored [21], [22], [23]. These techniques incur performance loss due to delays introduced by additional peripheral circuits to perform reordering, and error compensation. Software approaches, in contrast, rely solely on algorithmic solutions. Several existing investigations have explored software techniques for ReRAM-based PIM accelerators including thermal aware mapping, multi-objective optimization to balance accuracy with noise robustness, and retraining in the presence of stochastic noise [15], [16], [25]. Another set of software techniques specifically for CNN workloads introduce online tuning of mean and variance parameters of batch normalization (BN). These methods typically use a batch of training data or a batch of online test data to update the statistics, either with or without NAT [20], [40], [41], [10]. In addition to these, some approaches rely on frequent on-device training for noise mitigation. However, such retraining is impractical, as DNN training requires the computation of weight and activation gradients during backpropagation, which involves multiple write operations and NVM devices have limited write endurance [19] [26]. These critical drawbacks limit the applicability of existing training-based NAT techniques. Additionally, hybrid approaches that combine both hardware and software strategies have also been proposed to improve noise resilience [14], [21]. These include designs with temperature sensors with a corresponding software strategy to perform temperature-aware weight adjustments. Another work uses current mirror circuits to compensate for the deviation in the crossbar output due to thermal noise, in addition to thermal-aware mapping [21], [22]. However, all the above-mentioned approaches fail to provide consistent neural network inference accuracy across the entire operating temperature range. *The goal of this paper is to precisely fill this gap in the current state of knowledge.*

## 3 PROBLEM SETUP

NVM-based PIM devices are vulnerable to thermal noise, leading to degradation in DNN training/inference accuracy [16], [23]. This thermal sensitivity poses a major challenge for deploying DNNs on NVM-based PIM platforms [25]. Existing NAT techniques attempt to address this by adapting model weights to specific noise levels [15], [25]. However, once deployed on an NVM crossbar, these adapted weights deviate from their original values due to change in temperature, limiting their effectiveness. Frequent on-device retraining is also not a viable option due to limited write endurance of NVM devices [26]. In the following section, we discuss thermal noise modeling for NVM cells and outline the noise injection strategy used during re-training. This forms the basis for defining a hardware-aware retraining procedure to mitigate the effects of thermal noise.

### 3.1 Thermal Noise and Noise Injection Strategy

The correct modeling of thermal noise and an effective mitigation framework is critical for ensuring consistent DNN inference accuracy. Moreover, the impact of thermal noise varies across different NVM devices. Consequently, the same temperature induces different weight deviations across different devices. We consider the functional relationship between temperature and thermal noise characteristics for each device following existing literature [24], [34], [35], [42]. ReRAM devices store DNN weights as conductance (or resistance) states. As temperature increases, the OFF-state conductance of ReRAM cells rises exponentially, reducing the noise margin [24]. The OFF-state conductance exhibits exponential temperature dependence following the relationship shown below:

$$G_{off}(T) \propto e^{\left(-\alpha(\phi_o - \theta T)\right)} \tag{1}$$

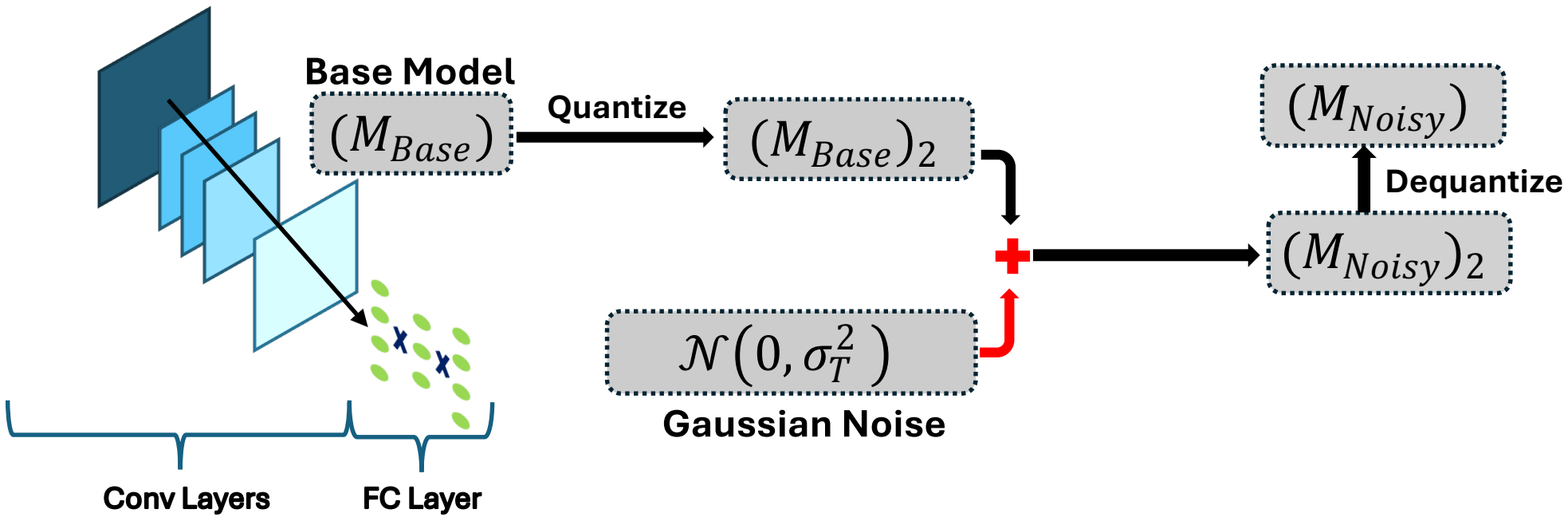

Figure 1: Overview of the thermal noise injection flow for noise-aware training. Thermal noise in NVM cells is modeled as Gaussian perturbations in weights and is injected into quantized DNN weights to emulate on-device behavior across temperatures. Here, $(M_{Base})_2$ represents the quantized model weights.

Where, $\alpha$ is a parameter related to the potential profile curvature at the constriction's bottleneck, $\theta$ is the temperature coefficient and $\phi_o$ is the zero-temperature barrier height, derived from the quantum point-contact (QPC) model [24]. On the other hand, PCM cells store weights as resistance states, where the amorphous state represents high resistance, while the crystalline state represents low resistance [34], [42]. Both amorphous and crystalline resistance states follow thermally activated transport with exponential temperature dependence, with electrical conductivity ($\sigma(T)$) following the relation:

$$\sigma(T) \propto e^{\left(-E_a/k_B T\right)} \tag{2}$$

Where, $E_a$ is the activation energy and $k_B$ is the Boltzmann's constant. In contrast, FeFET devices store weights as threshold voltage states determined by the polarization state of the ferroelectric material [35]. Unlike the exponential dependencies in ReRAM and PCM, the noise margin of FeFET devices, defined by the memory window between polarization states, shows linear temperature dependence through the Landau coefficient $\alpha_1$:

$$\alpha_1(T) = {\alpha_0(T_c - T)}/{T_c} \tag{3}$$

Where, $\alpha_0$ is a material parameter and $T_c$ is the Curie temperature, leading to linear degradation of noise margin [35]. We model weight variation considering these relationships of device characteristics with temperature, which form the foundation for modelling the thermal noise directly as a Gaussian distribution $\mathcal{N}(0, \sigma_T^2)$, where $\sigma$ represents the standard deviation of weights at temperature $T$ consistent with the existing work [24], [43], [44].

Next, we discuss how thermal noise is injected to the DNN model weights to imitate on-device behavior for NAT. Consider a base model $M_{Base}$ with an $i^{th}$ Convolutional or Linear layer ($L$) weight tensor represented by $[W]_i^L$, where each individual weight value falls within the quantization boundary $[-a, a]$. The $n$-bit quantization of these weights is expressed by:

$$[W]_{i\ clamp}^L = \min\left(max\left([W]_i^L, -a\right), a\right) \tag{4}$$

$$S = \frac{2^{n-1} - 1}{a} \tag{5}$$

$$[W]_{i\ Quantized}^{L} = round\ ([W]_{i\ clamp}^{L} \times S) \tag{6}$$

Equations (4)-(6) represent the standard signed quantization scheme. After quantization, the deviation in stored weights due to thermal noise at any temperature $T$, given by Gaussian noise tensor $\mathcal{N}(0, \sigma_T^2)$ is injected into the weights. The noise injection method can be represented by the formulation below:

$$[W]_{i\ \text{Noisy}}^{L} = [W]_{i\ Quantized}^{L} + \mathcal{N}(0, \sigma_T^2) \tag{7}$$

Following this approach, a standard de-quantization process is applied to obtain the noisy DNN model weights. This process is repeated for every convolution and linear layer, resulting in the noisy model $M_{Noisy}$. The entire process is illustrated in Fig. 1. The noisy model is then used for NAT, during which the model is re-trained multiple times at various discrete noise levels. To highlight the limitations of existing NAT techniques, Fig. 2 presents a conceptual view of a processing element (PE), tile, and core. Each PE has multiple tiles and other supporting units such as I/O block, Activation unit, embedded DRAM, control unit, and Special Function Units (SFUs). Each tile contains multiple cores with crossbars which store the DNN weights. In conventional NAT methods, the noise-aware parameters are loaded onto the NVM crossbars, where any change in device temperature leads to further weight deviation. This entire process is represented in Equations (8)-(9) and Fig. 2.

$$[W]_{base} \xrightarrow{Noise\ Injection} [W]_{Noisy} \xrightarrow{NAT} [W]_{NA} \tag{8}$$

$$[W]_{NA} \xrightarrow{Map\ to\ NVM\ crossbar} [W]_{NA}^{*} \tag{9}$$

Here, $[W]_{base}$ represents the noise free pre-trained model weights and $[W]_{Noisy}$ denotes the model weights after noise-injection. After NAT, we obtain $[W]_{NA}$, which represents the noise-aware weights adapted to the noise via backpropagation. These weights are expected to provide inference accuracy comparable to the noise-free model. However, when mapped onto crossbars, these weights deviate due to varying levels of thermal noise caused by changes in temperature during inference, resulting in the model using weights $[W]_{NA}^{*}$ instead. When the crossbar current is sampled by the ADC at the end of bit-lines (see Fig. 2), the obtained output is distorted, leading to degraded model inference accuracy.

Each device exhibits a different level of thermal sensitivity, as discussed above. Fig. 3 shows the inference accuracy achieved for the VGG16 model with the CIFAR-100 dataset on existing R-PIM, P-PIM, and F-PIM architectures when thermal noise is injected using Eq. 7 [4], [45], [46]. Among these, F-PIM shows the least reduction in accuracy due to its linear temperature dependence, as discussed in Section 3.1. Nevertheless, the noticeable degradation across all NVM architectures highlight the need for effective noise mitigation techniques to make the PIM architectures viable for real-world scenarios. Although NAT methodologies aims to ensure generalized robustness across the full operating temperature range, our subsequent evaluation demonstrates that existing methods fail to meet this objective.

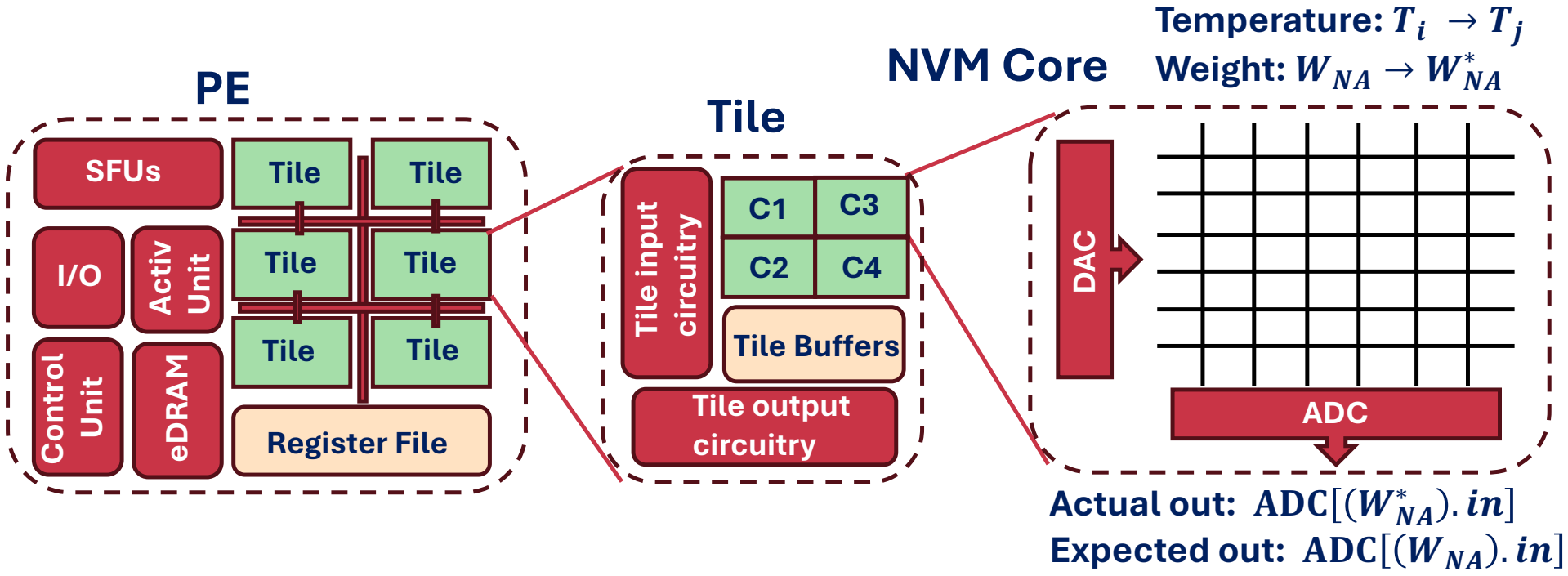

Figure 2 : A conceptual illustration of the PIM Processing Element (PE), Tile, and Core.

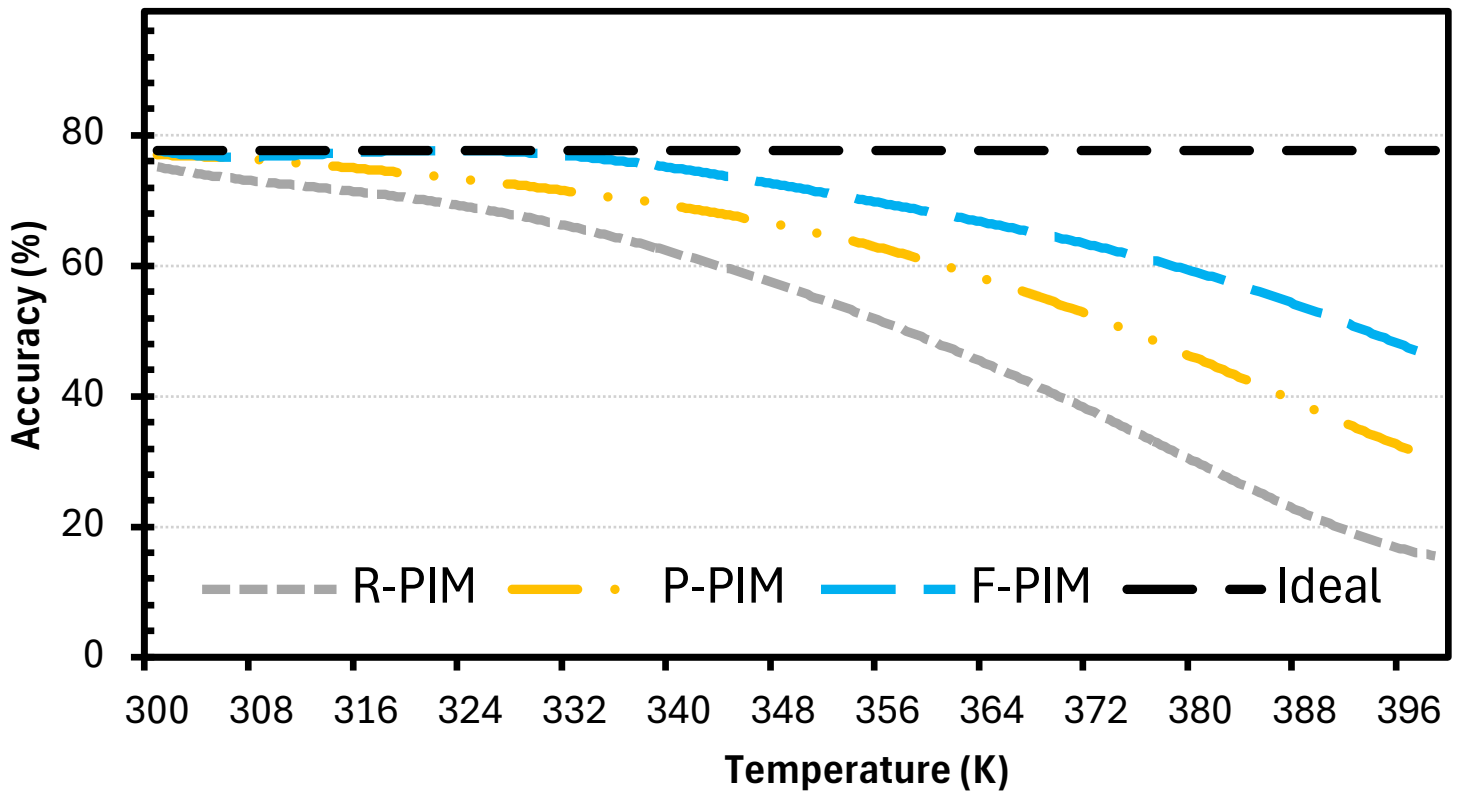

Figure 3: Inference accuracy achieved on R-PIM, P-PIM, and F-PIM architectures under thermal noise across the full operating temperature range ($300°K$- $400°K$) for the VGG16 model on the CIFAR-100 dataset.

## 4 THRIVE: NOISE-AWARE TRAINING METHODOLOGY

In this section, we propose a hardware and software co-design approach to mitigate thermal noise from NVM-based PIM systems and describe each part of the approach in detail.

### 4.1 ThRIve Hardware Framework

To overcome the limitations of existing methodologies and to mitigate thermal noise in NVM-based PIM systems, we propose an end-to-end solution utilizing LoRA based parameterization [27]. LoRA is a state-of-the-art fine-tuning technique for transformers that works by introducing two trainable low-rank matrices, A and B, into specific weight matrices of a pre-trained model, allowing adaptation to specific tasks without updating the original parameters. By constraining the weight update to a low-rank space, LoRA significantly reduces the number of trainable parameters and associated memory footprint than full fine-tuning while maintaining or improving the task performance [27]. Before NAT, we add these low-rank matrices to each neural layer that will be mapped onto the NVM crossbar. For example, we add LoRA parameters to the convolutional and linear layers in case of CNNs as shown in Fig. 4. This gives us the LoRA parameterized version of the base model. During the offline NAT phase, thermal noise is injected into the original model weights according to the procedure defined in Section 3. Afterwards, the model is re-trained in the presence of this noise. Here, the original weights are kept frozen after noise-injection and only the low-rank matrices get updated through backpropagation. Equation (10) shows the LoRA computation kernel:

$$Y = W_{NA}X = \left(W_{Noisy} + BA\right)X = W_{Noisy}X + BAX \tag{10}$$

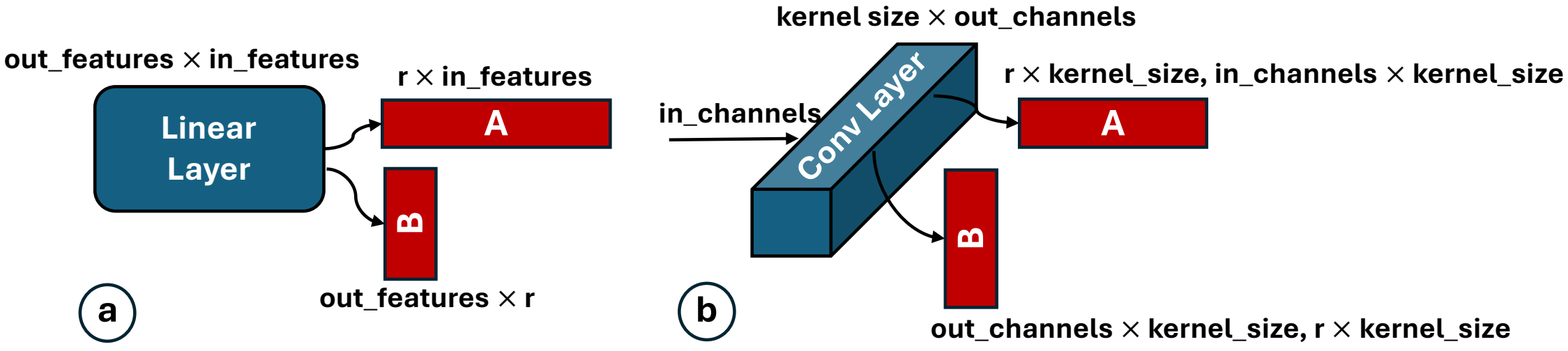

Figure 4: Low-rank adaptation for a) Linear layer and b) Convolutional layer using LoRA. Trainable low-rank matrices are added to the base model before NAT, while the original layers remain frozen during training.

Where, $Y$ is the output with input being $X$, and $W_{NA}$ are the weights after inclusion of noise-aware trained LoRA matrices ($B$ & $A$). The noise-aware weights $W_{NA}$ is the sum of the noise-injected weight ($W_{Noisy}$) which was kept frozen during NAT, and the product of low-rank matrices ($BA$). The dimensions of these low-rank matrices differ based on the type of layer, i.e. whether it is a convolution or linear layer. In the case of a linear layer, the dimension of LoRA matrices given the noise-injected weights $W_{Noisy}$ will be:

$$W_{Noisy} \in \mathbb{R}^{F_{out} \times F_{in}}$$

$$A \in \mathbb{R}^{(r \times F_{in})}, B \in \mathbb{R}^{(F_{out} \times r)} \tag{11}$$

Here, $F_{in}$ *and* $F_{out}$ represents the number of input features and output features for a Linear layer.

Conversely, for a convolutional layer with noisy weights $W_{Noisy} \in \mathbb{R}^{C_{out} \times C_{in} \times k \times k}$, the dimensions of corresponding LoRA matrices are given by:

$$A \in \mathbb{R}^{(r.k) \times (C_{in}.k)}, B \in \mathbb{R}^{\left(\frac{C_{out}}{groups}.k\right) \times (r.k)} \tag{12}$$

Here, $C_{in}$ and $C_{out}$ represent the number of input and output channels, $k$ is the kernel size, and $groups$ represents the independent channel partitions for a convolution layer. The rank "$r$" is the LoRA hyperparameter that is significantly smaller in dimension in comparison to the original parameter dimension ($r \ll F$ or $r \ll C$) in both the cases. This LoRA parameterized model is re-trained with different noise levels in the operating thermal range to make these trainable low-rank matrices noise-aware. However, if we use NVM-based PIM architectures and map the noise-aware LoRA weights onto these NVM crossbars, these weights will be further exposed to thermal noise, leading to deviation in their magnitude, as discussed in Section 3. This deviation reduces their effectiveness in mitigating thermal noise. For successful noise mitigation, these low-rank matrices must be stored in a noise-resilient PIM crossbar. To address this, we leverage existing NVM-CMOS heterogeneous PIM architectures, such as HyDE, HuNT, and HyperX, which integrate various NVM cell technologies such as ReRAM, PCM, and FeFET with CMOS-based SRAM [12], [13], [19]. In these architectures, we employ NVM to store the pre-trained model base weights, while the noise-aware LoRA parameters are stored in SRAM, a thermally robust CMOS-based memory.

During on-device deployment for inference tasks, the low-rank matrices $A$ and $B$ from the LoRA parameterized model are mapped to SRAM crossbar, while the model's base weights are mapped to NVM crossbars. The input is simultaneously fed into both SRAM and NVM compute units. Afterwards, the final activation is obtained from the summation of output activations from both these computations in accordance with the LoRA computational kernel presented in Equation (10). Fig. 5 illustrates the high-level flow of activations from the $i^{th}$ layer ($X_i^L$) to the $(i+1)^{th}$ layer ($X_{i+1}^L$) and shows how the noise-aware weights, $W_{NA}$ are obtained. This methodology ensures that the neural network maintains high accuracy and robustness, even in the presence of thermal fluctuations as even with noisy weights $W_{Noisy}$ (after mapping on the NVM crossbar), the model inference effectively happens using $W_{NA}$. Our proposed approach adds storage overhead with LoRA parameters on SRAM PIM. However, the resources required for these computations is significantly less than that of mapping all the model's base pre-trained weights to SRAM PIM. LoRA introduces only two low-rank matrices, as a result, the number of additional parameters and the corresponding MAC operations needed for the LoRA path ($B.A.X$ computation in Fig. 5) is substantially reduced compared to the full-rank weight matrix mapped on NVM crossbars. For instance, if we use $rank = 16$ for a convolutional layer with $C_{in} = 128$, $C_{out} = 256$ and with a kernel size of $3X3$, the LoRA path requires over ~82% fewer parameters and computations than the original convolution layer. Thus, storing and computing LoRA parameters on SRAM incurs minimal storage overhead. As illustrated in Fig. 5, both computations, i.e., using the model's base weights on NVM and the LoRA parameters on SRAM, can occur in parallel leading to negligible overhead in execution time. This makes ThRIve a practical and efficient solution for thermally-robust inference in heterogeneous PIM systems.

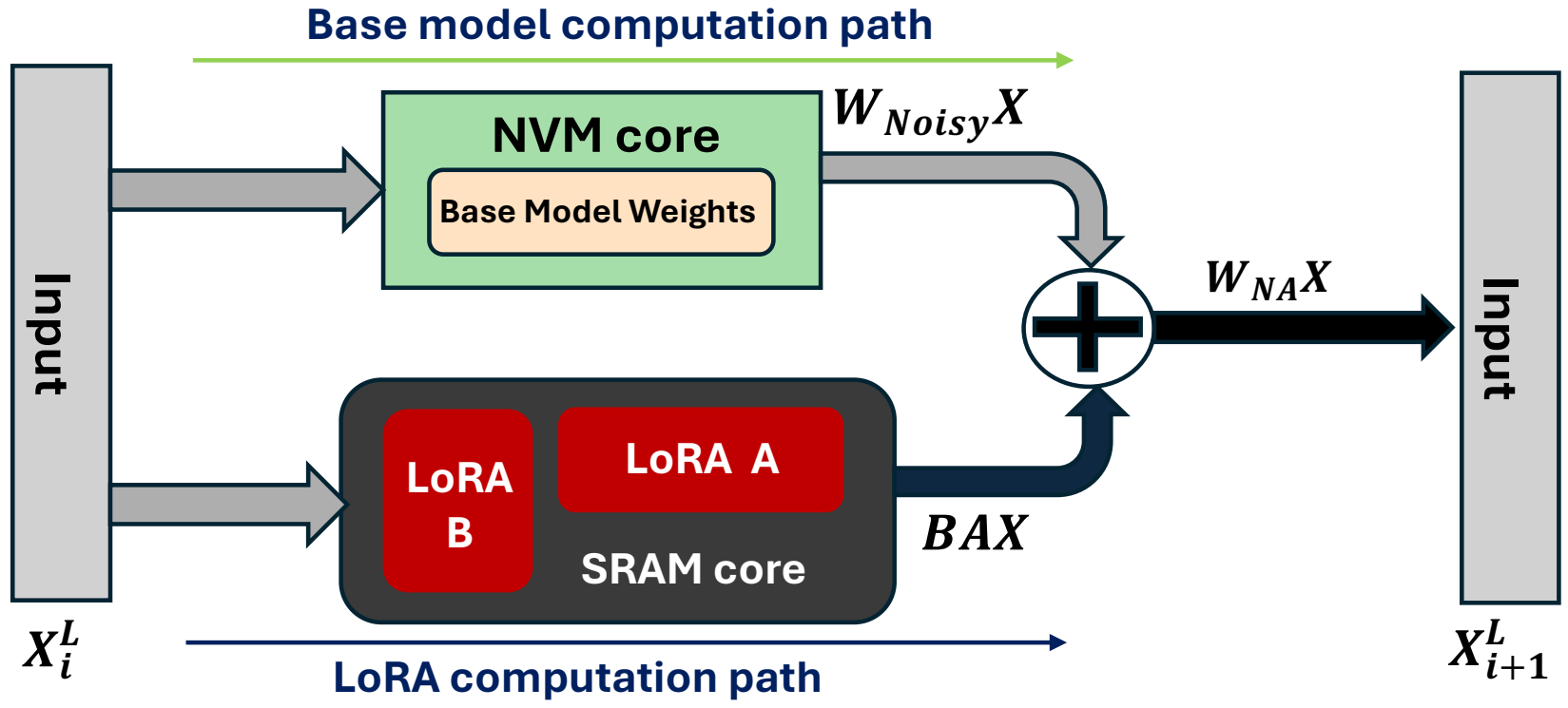

Figure 5: High-level illustration of activation flow in the ThRIve-enabled heterogeneous architecture from the $i^{th}$ layer $X_i^L$ to the $(i+1)^{th}$ layer $X_{i+1}^L$ during the post-deployment inference phase.

### 4.2 ThRIve Algorithmic Solution

**Motivation and Overview of our Approach:** A straightforward approach to make a model robust to noise is to train it separately for every possible noise level using the LoRA-based NAT method as discussed in the last section. However, this becomes impractical because the temperature range is continuous making the possible number of thermal noise levels to be very large. Moreover, each training run is costly. To handle this efficiently, we propose a novel approach that starts by initializing the given pre-trained model into its corresponding LoRA parametrization. Our key idea is to solve a min-max optimization problem where "max" is over the candidate noise values and "min" is over the LoRA parameters, and the goal is to find the LoRA parameters that minimizes prediction errors for worst-case noise values. Our iterative approach repeatedly solves the inner maximization problem (line 5 in Algorithm 1) using Bayesian optimization (BO) to find the worst noise followed by solving the outer minimization problem (line 2 in Algorithm 1) using LoRA based noise-aware training for the selected noise. We continue until convergence or maximum iterations whichever is earlier. Equation (13) express the *min-max optimization* problem.

$$\min_{\theta_{LoRA}} \max_{\eta \in \mathcal{N}} \mathcal{L}(f(X; \theta_{base} + \eta, \theta_{LoRA}), Y) \tag{13}$$

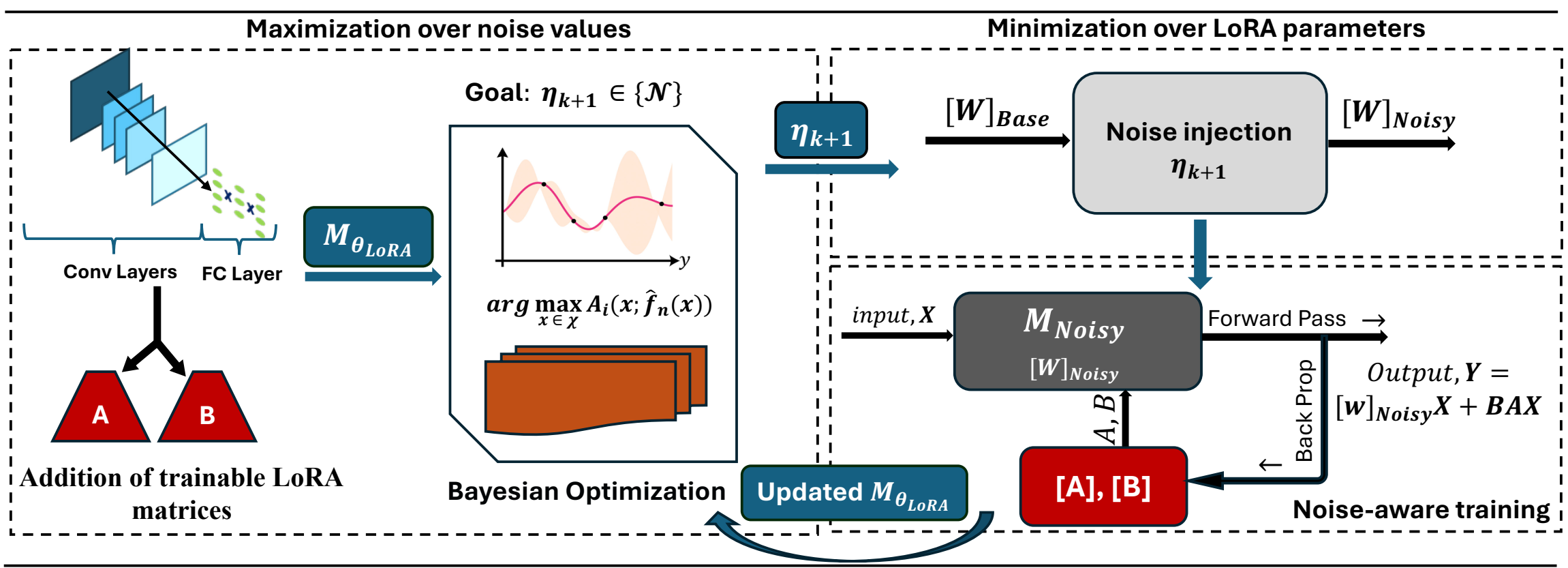

Figure 6: ThRIve algorithm overview. LoRA layers are added to the base model, which is then trained via a min-max loop: Bayesian Optimization selects worst-case noise, and LoRA parameters are trained during NAT to minimize the loss. This yields a model robust to thermal noise in the entire operating temperature range.

Here, $\theta_{LoRA}$ are the trainable low-rank parameters, $\theta_{base}$ are the original model parameters, $\eta$ represents the noise in the noise space $\mathcal{N}$ and $X, Y$ are the input and target output for the model.

Now, we describe the three main algorithmic steps of our methodology: initialization, selection of worst noise using BO, and LoRA based training with the selected noise.

**1. Initialization:** In this phase, each layer of the model that will be mapped on the NVM-based PIM is added with two trainable LoRA matrices as discussed in Section 4.1. After which the original weights of the model are frozen leaving only the LoRA weights to be trainable.

**2. Bayesian Optimization Phase:** BO is a sample-efficient method to adaptively and efficiently search a given input space (search space of noise values based on operating temperature range) to optimize expensive-to-evaluate objective function (prediction loss of a given deep model in the presence of noise). The key parts of BO for data-driven decision-making are: 1) a **surrogate model** that captures our beliefs, based on past objective function evaluations, about the input-output relationship; and 2) an **acquisition function** that scores each input according to the utility of querying the objective function on it next. We employ a Gaussian process (GP) surrogate model with Radial Basis Function kernel as our surrogate model. The acquisition function scores the utility of evaluating the next input with the expensive objective function. Here, "utility" is defined in terms of our goal of finding the optimal input (worst noise) with the minimum number of objective function evaluation queries (prediction loss evaluation for different noise values).

The decision of which input to evaluate next is made by maximizing the acquisition function. Importantly, the acquisition function is cheap to evaluate. There are many acquisition functions in the literature including expected improvement (EI), probability of improvement (PI), upper confidence bound (UCB), and lower confidence bound (LCB) [47]. However, their effectiveness varies across different tasks [48], [49]. Motivated by this observation, we consider a library of acquisition functions and adaptively select one of them in each BO iteration based on their past performance. We also include *random sampling* into our library of acquisition functions to avoid the potential risk of local optima. The output of the BO phase is an input from the noise space on which the model will be re-trained in the next phase.

**3. LoRA based Noise Aware Training Phase:** In this phase, the input noise level predicted by the BO phase is injected into the model. After which the model's original layers are frozen and the LoRA layers are kept trainable, and

**Algorithm 1. ThRIve Algorithm**

**Input:** A pre-trained deep model $M_{\theta_{base}}$, Noise search space $\mathcal{N}$, Library of BO acquisition functions $A_Q$, Convergence criteria
**Output:** A thermal noise-robust deep model $M_{NA}$

```
    /* Initialization */
1:  M_θLoRA ← ApplyLoRA (M_θbase) /* LoRA parameterization: Freeze base weights and initialize LoRA parameters */
2:  InitGP(η, L(η))               /* Fit Gaussian Process on initial noise-loss pairs */
    /*minimization loop*/
3:  Repeat:
4:      Set worst-case noise η_worst ∈ N
        /*maximization loop*/
5:      for each iteration i = 1 to Z  /* Here Z is the no. of acquisition functions in the library A_Q */
6:          Select acquisition function a_i ∈ A_Q
7:          Obtain the next noise candidate η_i ∈ N from a_i
8:          Evaluate prediction loss L(η_i) of the current model using η_i
9:          Update η_worst ← η_i, if loss L(η_i) > L(η_worst)
10:     end for
11:     Train M_θLoRA using the selected worst-case noise η_worst
12:     Evaluate L(η_i) for all {η_i}_{i=1}^Z and Update surrogate model with {(η_i, L(η_i))}_{i=1}^Z
13: Until convergence or maximum iterations
14: Return M_NA, the noise aware model.
```

the model is re-trained. This updates the LoRA weights via backpropagation resulting in a noise-aware LoRA parametrized model.

The total computational time of our BO-NAT methodology follows the relationship:

$$t_{computation} = t_{initial} + B \times \left(A \times t_{epoch} + t_{GP_{OPS}}\right) \tag{14}$$

Where, $t_{initial}$ represents a one-time initial exploration cost for building the surrogate model, $B$ represents the number of BO iterations required for convergence, $A$ denotes the epochs per training session in the presence of selected noise value, $t_{epoch}$ corresponds to the wall-clock time per training epoch and $t_{GP_{OPS}}$ is time taken for BO surrogate model operations (Lines 5-10 and 12, Algorithm 1) . Here, the training under noise (Line 11, Algorithm R1) represents the most computationally expensive operation since it involves training the LoRA parameters through backpropagation across multiple epochs. In contrast, the BO surrogate model operations including acquisition function selection and noise candidate evaluation over a batch requires significantly less computational time compared to the neural network training phase. The full algorithmic flow is presented in Fig. 6 and Algorithm 1 shows a high-level pseudocode for our noise-aware training methodology based on Bayesian Optimization.

## 5 EXPERIMENTAL RESULTS

In this section, we present a detailed experimental evaluation of the proposed ThRIve methodology. We begin by outlining the experimental setup used for the performance evaluation. Next, we assess the limitations of existing noise mitigation approaches to establish a baseline for comparison. Subsequently, we present the overall performance of ThRIve in terms of accuracy and energy efficiency by applying it to existing NVM-CMOS heterogeneous PIM architectures. Finally, we evaluate the adaptability of ThRIve across various datasets.

Table 1: PIM architecture modelling specifications (*$F$ is the minimum feature size [12], [13])

| NVM Type | Cell Type, Resolution | Crossbar Configuration | Cell Area |
|---|---|---|---|
| ReRAM | 1T-1R, 2-bit/cell | $128 \times 128$ | $4F^2$ |
| PCM | 1T-1R, 2-bit/cell | $256 \times 256$ | $4F^2$ |
| FeFET | 1T, 2-bit/cell | $256 \times 256$ | $6F^2$ |
| SRAM | 6T, 1-bit/cell | $256 \times 256$ | $150F^2$ |

Table 2: List of CNN workloads along with their number of parameters (in Millions)

| | VGG16 | ResNet18 | MobileNetv3_Small |
|---|---|---|---|
| CIFAR10 | 15.11 | 11.17 | 1.53 |
| CIFAR100 | 15.13 | 11.22 | 1.62 |
| TinyImageNet | 17.44 | 11.27 | 1.72 |

## 5.1 Experimental Setup

### *5.1.1 PIM Architectures and Tools*

In this work, we examine existing PIM architectures to highlight their limitations against thermal noise. For this analysis, we consider three NVM-based PIM architectures: R-PIM, P-PIM, and F-PIM. R-PIM is a ReRAM-based PIM architecture that performs energy-efficient MVM using ReRAM crossbars and supports both on-device training/inference [4]. It comprises of 64 PEs, each containing 96 crossbar arrays of size 128×128 [4]. P-PIM, on the other hand, uses PCM-based crossbar to store DNN weights with each crossbar of dimension 256×256 [45]. Since the crossbar size in P-PIM is four times larger than that in R-PIM, we consider a 32 PE system with 48 crossbars in each PE. F-PIM employs FeFET-based analog synapses, that enables fast and energy-efficient on-chip training of DNNs [46]. It comprises of 32 PEs with 6 tiles/PE where each tile has 8 crossbars of size 256×256 [46]. Additionally, we consider three existing NVM-CMOS heterogeneous PIM architectures: HyperX, HyDE, and HuNT. HyperX is a hybrid ReRAM-SRAM system. In this architecture, the DNN is partitioned such that layers with fixed (non-trainable) weights are mapped to the ReRAM blocks, while layers with trainable weights are mapped to the SRAM block. It aims to mitigate ReRAM non-idealities through on-device retraining, where the fixed weights remain unchanged, and only the trainable weights are updated during the training process [19]. We consider a 64 PE system in our evaluation without loss of generality. Subsequently, the HyDE framework proposes a hybrid-device search methodology that selects the optimal device type out of PCM, FeFET, or SRAM, for each DNN layer to balance inference accuracy, area, and energy under device-level non-idealities and comprises of 64 PEs [12]. HuNT on the other hand, integrates SRAM, ReRAM, and FeFET PIM devices to enable high performance and energy-efficient DNN training. It employs a total of 64 PEs, with 4 tiles/PE. Each tile has a different number of crossbars/arrays based on the cell type (96 for ReRAM, 48 for PCM and 9 SRAM arrays). HuNT considers several factors such as power, area, endurance, and accuracy when mapping each DNN layer to a corresponding PIM device [13].

In the hardware evaluation, area, energy, and latency of the ReRAM, PCM, FeFET and SRAM devices and their associated peripheral circuits such as ADC, sense-amps (S/A), DACs, buffers, column-/row-decoders, nonlinear activation units (ReLU) were modeled via NeuroSim [50], [51]. NeuroSim incorporates cycle-accurate analytical tools (NVSim & CACTI) to evaluate the compute area, latency, and power for each PIM configuration. Table 1 presents the bit/cell resolution, crossbar configuration and cell area of each PIM device used in the evaluation. Following prior device-level studies, thermal noise in ReRAM, PCM, and FeFET-based memory cells is modeled as a zero-mean Gaussian perturbation added to the ideal stored weight values (as shown in Equation 7) [15], [24], [43], [44]. This noise model captures the distinct temperature-dependent behaviors of different NVM devices as discussed in Section 3.1. These device-specific characteristics are incorporated in the noise model through tunable deviation coefficients, allowing Gaussian noise to mimic thermal noise behavior across different device types and operating temperatures, consistent with prior works [15], [24], [43], [44]. This enables evaluation of inference accuracy over the operating temperature range while accounting for the unique thermal sensitivity of each memory technology.

We employ a PyTorch-based wrapper, integrated with NeuroSim, to simulate the impact of temperature-induced thermal noise on weights stored in NVM devices, where model weights are represented in 16-bit fixed-point precision. Since a single NVM cell can store multiple bits of data, each quantized weight is decomposed into multiple sub-components representing individual memory cells within a crossbar structure [4], [50]. Throughout this work, we assume a 2-bit/cell storage capacity for NVM devices, as illustrated in Table I, and incorporate this multi-bit resolution in our noise injection framework to accurately simulate the impact of thermal noise. The operating temperature range for evaluating the PIM system is set between $300°K$ and $400°K$, which reflects the typical thermal conditions under which these memory devices operate [21], [23], [25]. During NAT, the noise values are adaptively selected from the noise search space (i.e., thermal noise corresponding to the $300°K$-$400°K$ operating temperature range) via the ThRIve framework and is injected into the model weights. For inference evaluation under noise, the temperature range is discretized into 50 uniformly separated points, with corresponding thermal noise injected into the model one by one to simulate the thermal variations. We then calculate the generalized accuracy as the average accuracy over all points, and the range deviation as the difference between the maximum and minimum accuracy values relative to the mean.

Further, we initialize the LoRA matrices using the standard approach, where matrix A is initialized with Kaiming-uniform distribution and matrix B is initialized to zeros according to the original literature [27]. This ensures that the product BA = 0 at initialization, maintaining the pretrained model's behavior at the start of fine-tuning [27]. Note that alternative initialization strategies can also be employed without affecting the framework's applicability or generalization. The maximum BO iterations is set to 60 whereas, the maximum training epochs per BO iteration is capped at 10. These settings enable ThRIve to generalize across diverse noise conditions without overfitting to a limited subset and minimizes the overall computational cost of constructing thermally robust models. To report model accuracy using ThRIve, each experiment is repeated 10 times, and the results are presented as the average across these runs. The experiments are executed on an Nvidia A40 GPU with 46GB of memory with Linux operating system. We employ the same GPU platform to perform the NAT step within the ThRIve software algorithm. This NAT procedure is a one-time operation executed on the GPU for each dataset as a pre-processing step. The cost of performing NAT is thus amortized over multiple inference runs on the accelerator.

#### *5.1.2 CNN models and Datasets*

We assess the performance of the ThRIve NAT methodology considering three CNN models namely: VGG16, ResNet18, and MobileNetv3_Small (MobileNetv3_S) with three diverse datasets CIFAR-10, CIFAR-100, and TinyImageNet [52], [53], [54], [55], [56]. Table 2 presents different CNNs, corresponding datasets, and the number of model parameters. Note that, the CNNs considered in our evaluations consist of linear (VGG), residual (ResNet), as well as factorized and attention-enhanced (MobileNetV3_Small) connections. Moreover, all the DNNs comprise of both fully connected and convolution layers. The evaluation with various CNNs demonstrates the broad applicability of ThRIve.

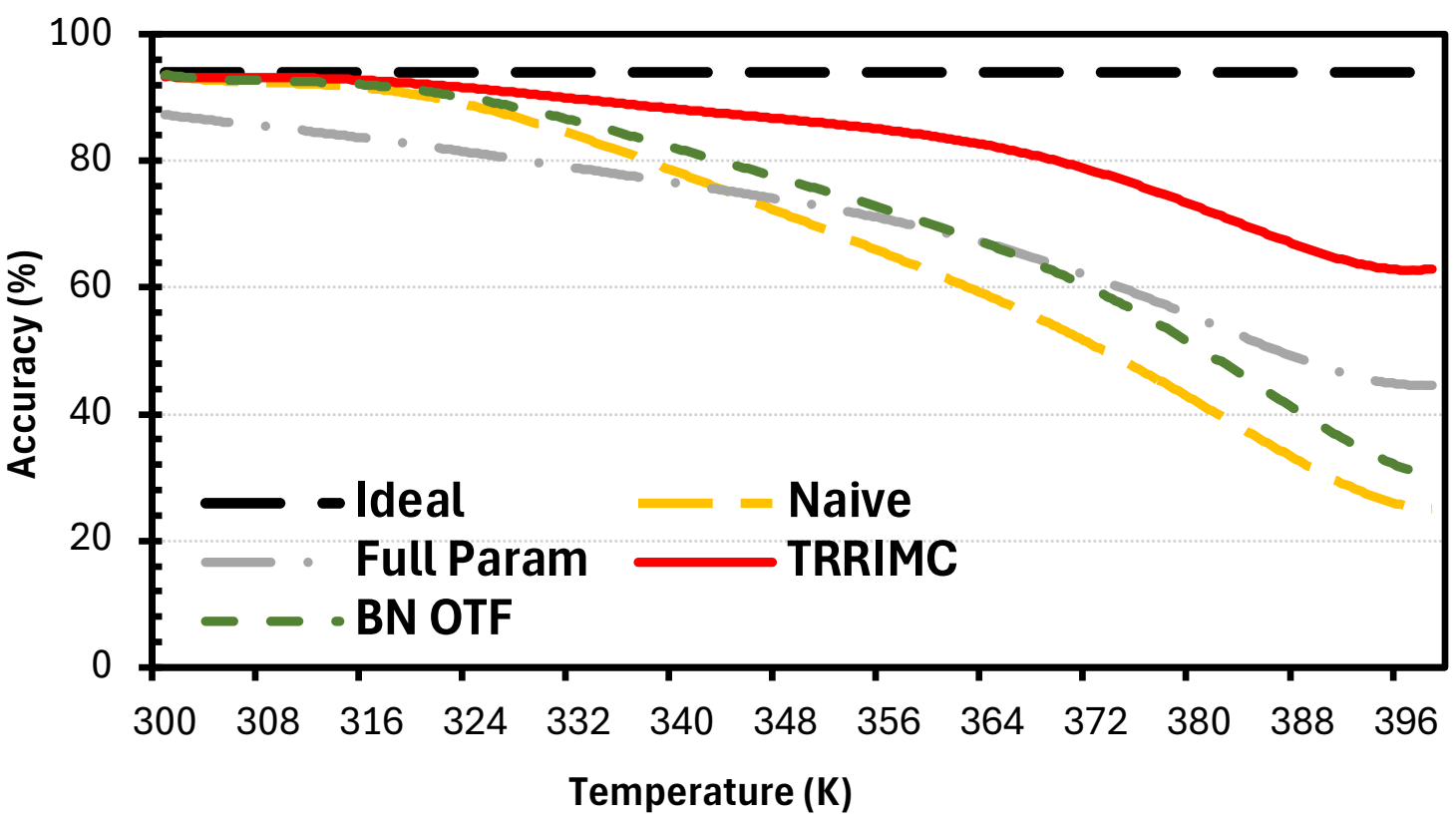

Figure 7: Generalized inference accuracy of ResNet18 on the CIFAR-10 dataset under thermal noise across the temperature range ($300°K$- $400°K$), evaluated on the R-PIM architecture. Various NAT techniques: Full Param, TRRIMC, and BN OTF are compared against a naïve baseline and an ideal case.

### 5.2 Limitations of Existing Noise Mitigation Approaches

In this subsection, we highlight the limitations of existing noise mitigation methodologies and explain why NAT approaches based solely on NVM-based PIM architectures are ineffective in mitigating the effects of noise. For this analysis, we consider state-of-the-art offline NAT techniques such as TRRIMC and NIA, which attempt to handle thermal noise by retraining all the base model parameters (Full Param) [15], [25]. We also include methodologies specifically designed for CNN workloads (BN OTF) which update BN statistics (mean & variance) online using a batch of training/test data. These updates can be done with or without offline NAT of BN parameters (shift & scale) [10], [20], [40]. Additionally, we examine techniques focused on reducing the peak temperature $T_{Peak}$ through thermal-aware mapping and other strategies [16], [21]. However, it is important to note that the drop in accuracy in such cases remains similar to the naïve approach (i.e., model is deployed on-device without any noise-aware training). The only difference is that the accuracy only drops until the system reaches the $T_{Peak}$. To evaluate these approaches, we employ the ResNet18 model with the CIFAR10 dataset noting that we see similar trends on other experimental configurations.

In our experiments, we first perform full parameter NAT, by injecting different noise levels into the model. This model is then deployed on a ReRAM crossbar-based architecture, R-PIM as a representative example, and evaluated across the full operating temperature range to simulate post-deployment thermal noise effects. For instance, in the TRRIMC method, thermal noise corresponding to two temperature values $(300°K, 310°K)$ are used for full parameter NAT using PKD, while thermal noise corresponding to three temperature values $(330°K, 360°K, 400°K)$ are used for BNA [25]. The resulting model is evaluated across the entire $300°K$- $400°K$ range to compute its generalized accuracy. We define generalized accuracy as the model's inference accuracy at unseen temperatures across the full operating range. This metric captures the model's ability to maintain robust performance under varying thermal noise conditions without requiring retraining for each specific temperature. For BN OTF, BN statistics are updated online using a batch of images [10], [20], [40].

Fig. 7 shows the generalized accuracy for all the techniques discussed above, including a naïve baseline where the model is deployed on-device without any NAT. While NAT techniques outperform the naïve approach, they still fail to provide consistent accuracy across the full temperature range. This is because the noise-aware model weights are mapped onto noisy ReRAM crossbars after NAT, where thermal noise causes further deviation in their magnitude. This reduces their ability to mitigate thermal noise effectively. Additionally, the continuous nature of device temperature makes these discrete NAT techniques even less effective. These methods rely on sequential/continual learning at discrete noise levels. This involves training on gradually increasing thermal noise. As the model progresses from one noise level to the next, it suffers from catastrophic forgetting, where performance on earlier noise conditions degrades. This effect becomes more pronounced as the number of training points increases, a known challenge in sequential learning frameworks [57]. Together, these limitations make existing NAT techniques ineffective. This is clearly reflected in the generalized accuracy achieved by the Full Param approach, which is trained on five uniformly spaced thermal noise levels. In this case, the model forgets about earlier noise levels, resulting in even worse accuracy than naïve baseline at lower temperatures. The effect of catastrophic forgetting is less pronounced in other techniques. TRRIMC mitigates it by using knowledge distillation (PKD) while transitioning between noise levels, and BN OTF achieves better noise recovery by updating BN statistics online. These issues highlight the need for heterogeneous architectures that combine NVM-based PIM with a thermally resilient counterpart, such as SRAM-based PIM, to reliably store the noise-aware parameters. A more effective NAT framework is required to address the problem of selecting and ordering thermal noise values during training.

### 5.3 LoRA Matrix Rank Analysis for ThRIve

In this section, we first investigate the impact of different LoRA ranks on the generalized accuracy of the model. Next, we discuss how different LoRA ranks affect the EDP. To calculate the rank for LoRA matrices, we follow an adaptive strategy that scales the rank based on the dimensions of each layer. Equation (15) and (16) show how the rank is computed for linear and convolution layers, respectively:

$$rank_{linear} = \lfloor F_{out} * P \rfloor \tag{15}$$

$$rank_{convolutional} = \lfloor C_{out} * P \rfloor \tag{16}$$

Here, $F_{out}$ denotes the number of output features in a linear layer, $C_{out}$ denotes the number of output channels in a convolutional layer, and $P$ denotes the rank proportion. This adaptive formulation allows the rank to scale automatically according to the size of each layer, ensuring efficient and size-aware parameterization across the network.

Fig. 8 shows the comparison of generalized accuracy for the three types of NVM devices considered in this work across various values of the rank proportion $P$. For this analysis, we assume the original model weights are mapped onto NVM cells, while the LoRA parameters are mapped onto thermally resilient CMOS-based hardware, as proposed earlier. We employ device-specific thermal noise modeling, discussed in Section 3.1, to reflect the temperature-dependent behavior of different NVMs. We employ the ResNet18 model with the CIFAR-100 dataset as a representative example. Further, to evaluate the model's ability to maintain consistent accuracy over the full operating temperature range, we employ two metrics: the mean of the generalized accuracy and the range deviation. The range deviation measures the difference between the maximum and minimum accuracy values relative to the mean. The figure reports both metrics across the entire operating temperature range for ReRAM, PCM, and FeFET-based PIM systems. From the plot, we observe that when original model weights are mapped onto ReRAM-based PIM and the $P$ value is smaller than 1/12, the generalized accuracy is significantly worse, showing a larger range deviation and lower mean accuracy. In contrast, for $P$ values greater than 1/12, accuracy plateaus and shows little to no improvement. However, the number of LoRA parameters increases as a higher rank proportion (i.e., higher rank) results in larger LoRA matrices.

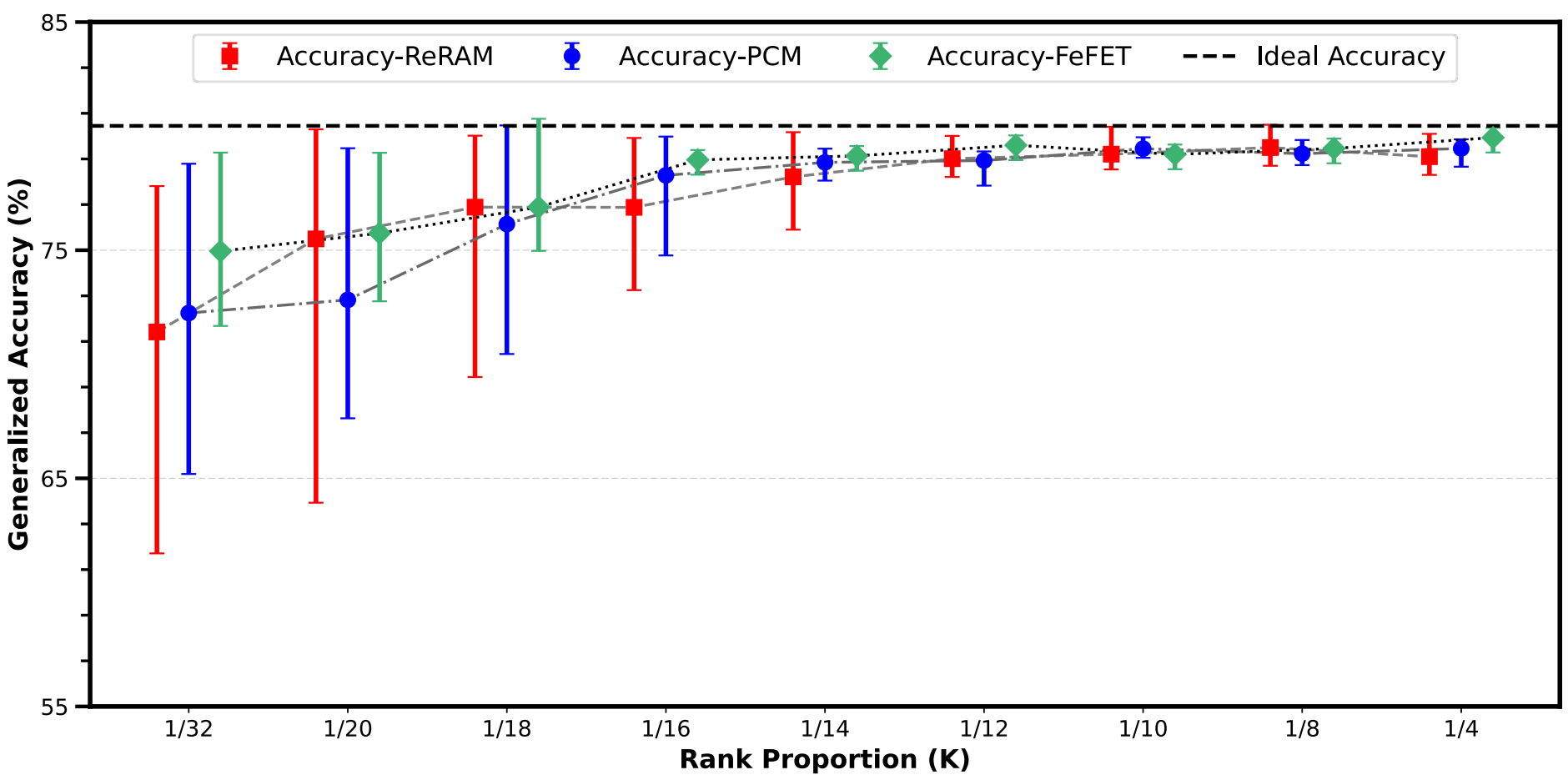

Figure 8: Comparison of generalized accuracy across varying LoRA rank proportions $P$. Results are reported for the ResNet18 model with the CIFAR-100 dataset. Ideal accuracy serves as the noise-free accuracy reference. Error bars indicate accuracy range deviation across the full operating temperature range.

Based on this ablation study, a rank proportion of 1/12 presents a good balance between generalized accuracy and the number of additional parameters for ReRAM-based PIM. As discussed in Section 3.1 and shown in Fig. 3, different NVM device types: ReRAM, PCM, and FeFET, exhibit varying levels of thermal sensitivity, resulting in different degrees of accuracy degradation. Consequently, similar accuracy can be achieved with lower $P$ values (i.e., smaller ranks) for devices that are less sensitive to thermal noise than ReRAM. For instance, PCM-based PIM maintains high mean accuracy with minimal deviation at $P = 1/14$, while FeFET-based PIM achieves near-ideal accuracy with low deviation at $P = 1/16$.

The LoRA parameters, mapped to the CMOS-based hardware, necessitate additional computation during the model inference. We now quantify the overhead in EDP due to using LoRA parameters with different rank proportion ($P$). Fig. 9 illustrates the corresponding impact of the rank proportion on EDP for ResNet18 model on CIFAR-100, evaluated on the HyperX architecture (ReRAM-SRAM). Here, EDP values are normalized to the architecture's baseline EDP without LoRA-based computation. From Fig. 9, we observe a clear trend that the EDP increases consistently with higher rank proportion values. Hence, a lower rank proportion (or lower rank) is desirable, as it reduces the number of additional parameters and, consequently, the EDP overhead. In this work, we utilize existing NVM-CMOS heterogeneous PIM architectures such as HyDE, HyperX and HuNT, which integrate different types of NVM cell technologies with CMOS-based SRAM. Without loss of generality, we assign a fixed rank proportion ($P$) of 1/12 for weights stored on ReRAM-based PIM, 1/14 for PCM-based PIM, and 1/16 for FeFET-based PIM to maintain a balanced trade-off between generalized accuracy and the number of LoRA parameters and hence EDP overhead as illustrated in Fig. 8 and 9.

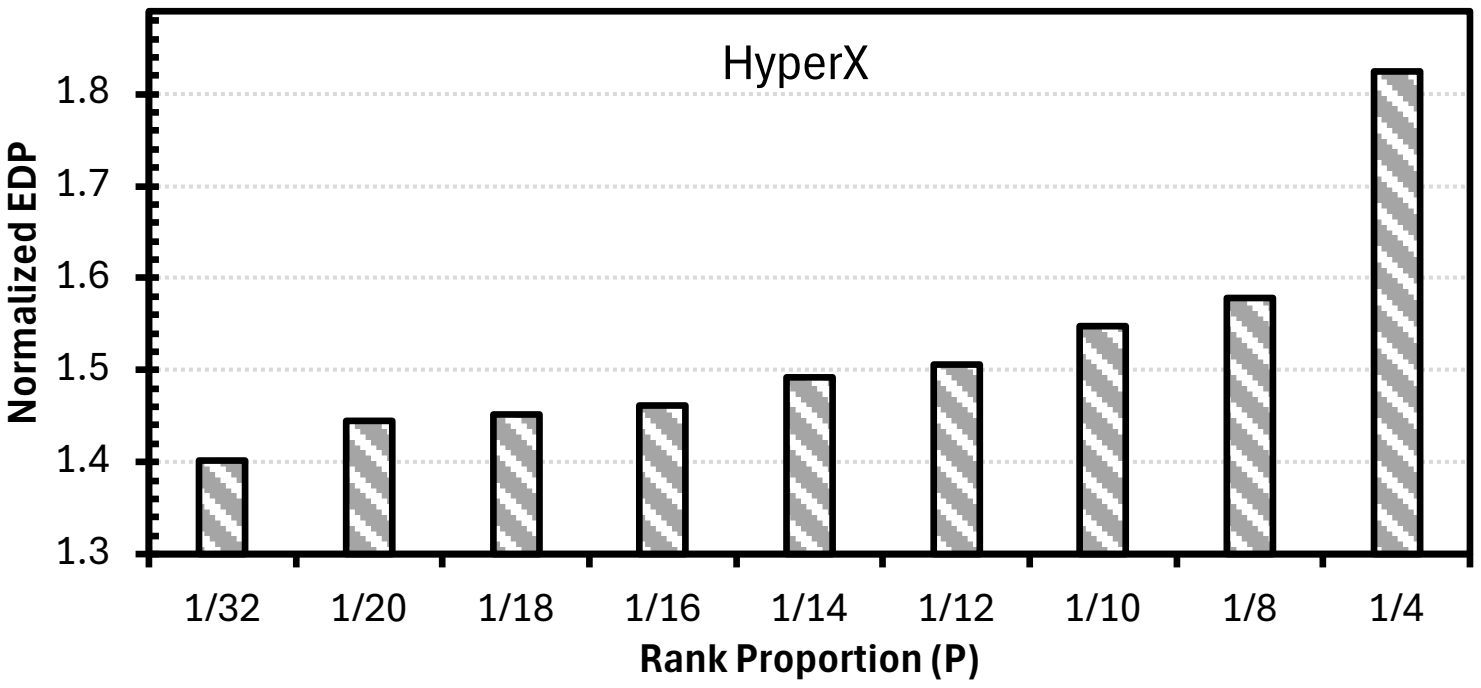

Figure 9: Comparison of normalized EDP for HyperX across varying LoRA rank proportions $P$. Results are reported for the ResNet18 model with the CIFAR-100 dataset.

### 5.4 Conventional Training vs. Bayesian Optimization based Noise-Aware Training

In this section, we present a comparison between the conventional training approach used in existing NAT methodologies and ThRIve's BO-based training strategy. Conventional approaches rely on sequential learning at discrete noise levels, where the model is trained on gradually increasing thermal noise, as discussed in Section 5.2. However, as training progresses, model's ability to recover accuracy on earlier noise conditions degrades. This approach has two key limitations (i) recovered accuracy is highly sensitive to the order of thermal noise levels used during training (ii) the continuous temperature range creates a large noise space, making representative point selection challenging.

For this analysis, we employ the HyperX architecture (ReRAM-SRAM heterogeneity) using the VGG16 model with the CIFAR-10 dataset. Fig. 10(a) presents the generalized accuracy for three different training orders of five noise levels (N=5), each initialized with a different random seed. We observe that recovery performance is highly order dependent, where Order-A and Order-B show better accuracy in the lower temperature range but suffer from a decline at higher temperatures. In contrast, the sequential order clearly exhibits the catastrophic forgetting discussed in Section 5.3, where the model performs well on the most recently trained noise levels but poorly on earlier ones. Therefore, selecting a suitable order and choice of noise values for training is a challenging problem. Fig. 10(b) highlights the point selection problem, where $N$ represents the number of training points sampled uniformly (as an example) over the entire temperature range (from the full noise sample space). The figure shows that for a certain order of points (kept same for all the cases) increasing the number of points improves recovery, with the generalized accuracy curve for $N$=15 being closest to the ideal accuracy line. However, selecting training points from a continuous thermal noise space is challenging. It is difficult to determine how many points to choose, and exactly which points should be selected. Additionally, deciding a suitable order for these points requires extensive experimental evaluation. This makes the overall process inefficient and computationally expensive even for offline execution. Moreover, this exhaustive search does not guarantee generalization across the entire operating temperature range.

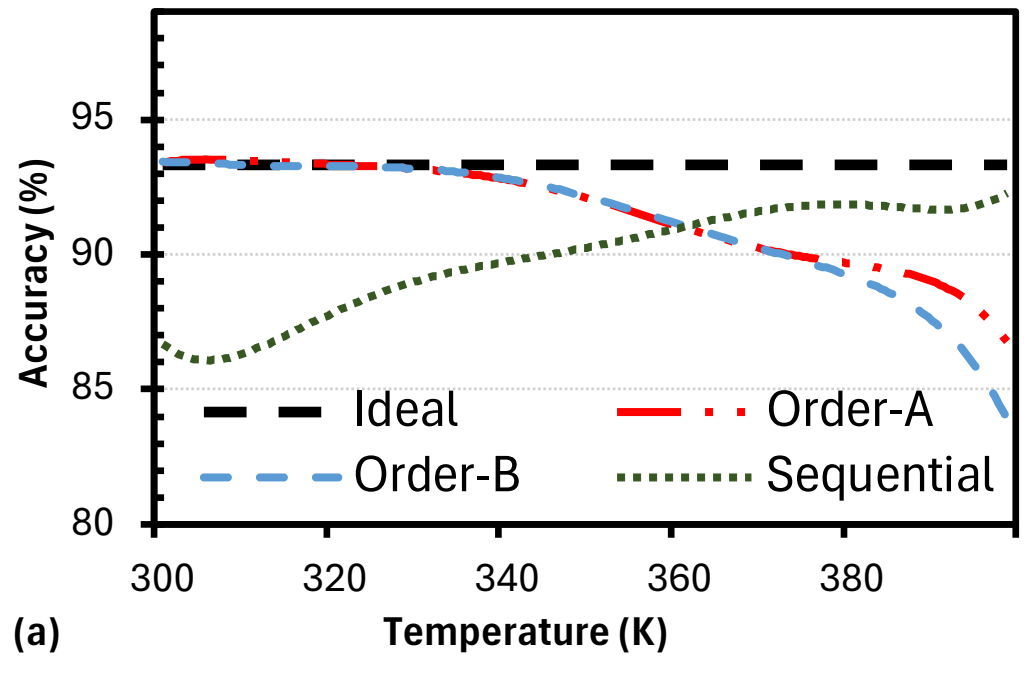

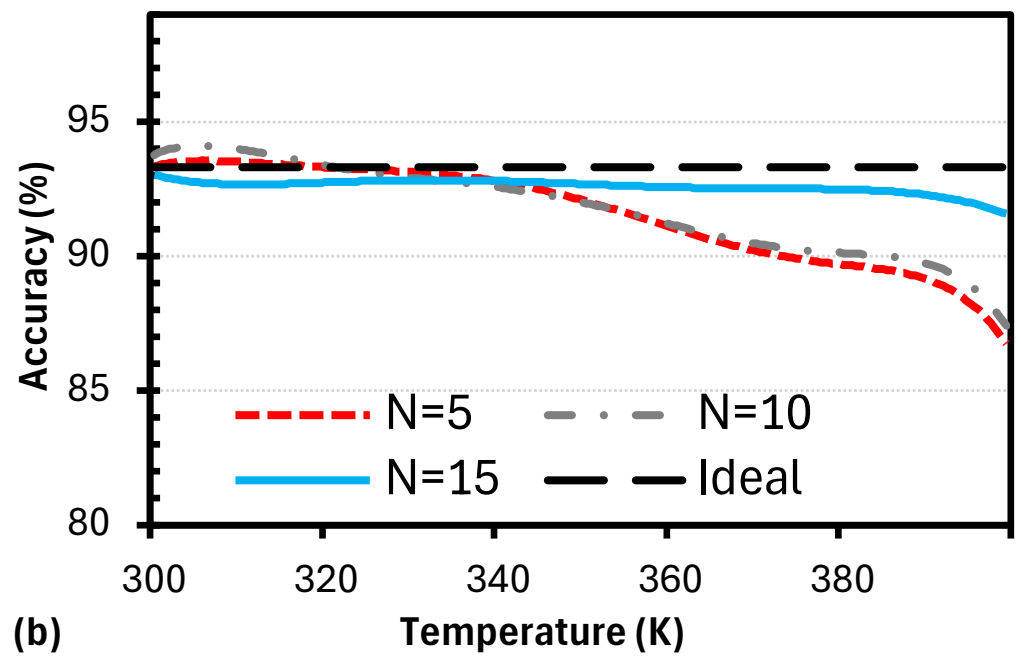

Figure 10: Generalized accuracy analysis of conventional NAT strategies on the HyperX architecture using VGG16 with the CIFAR-10 dataset. (a) Sensitivity to training order for $N = 5$, showing that recovered accuracy varies significantly with the order of noise levels used for NAT. (b) Effect of the number of uniformly sampled thermal noise levels ($N$) on accuracy recovery.

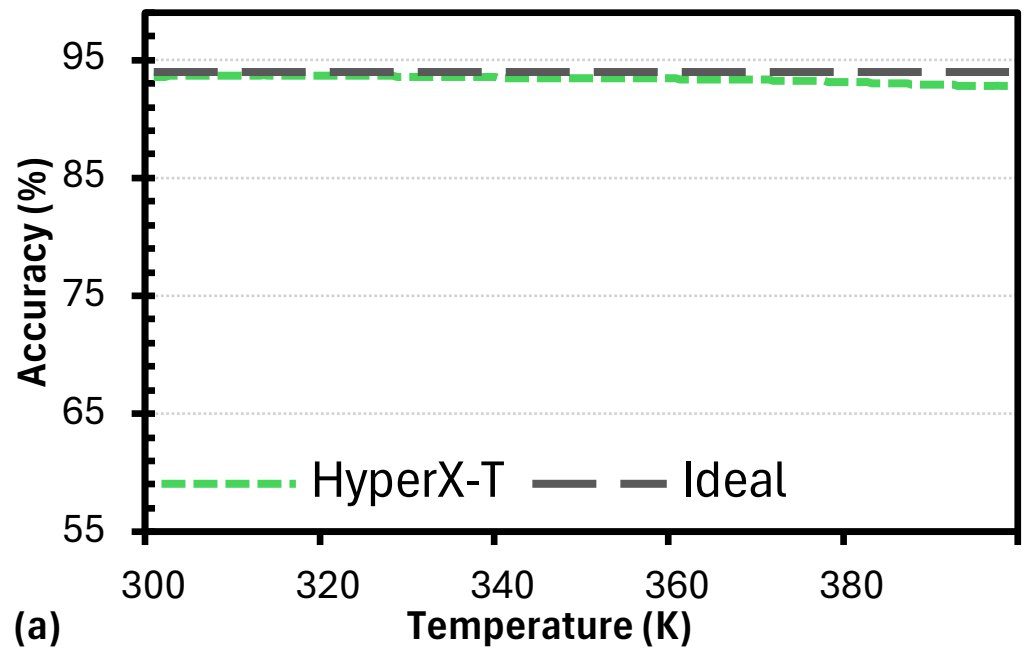

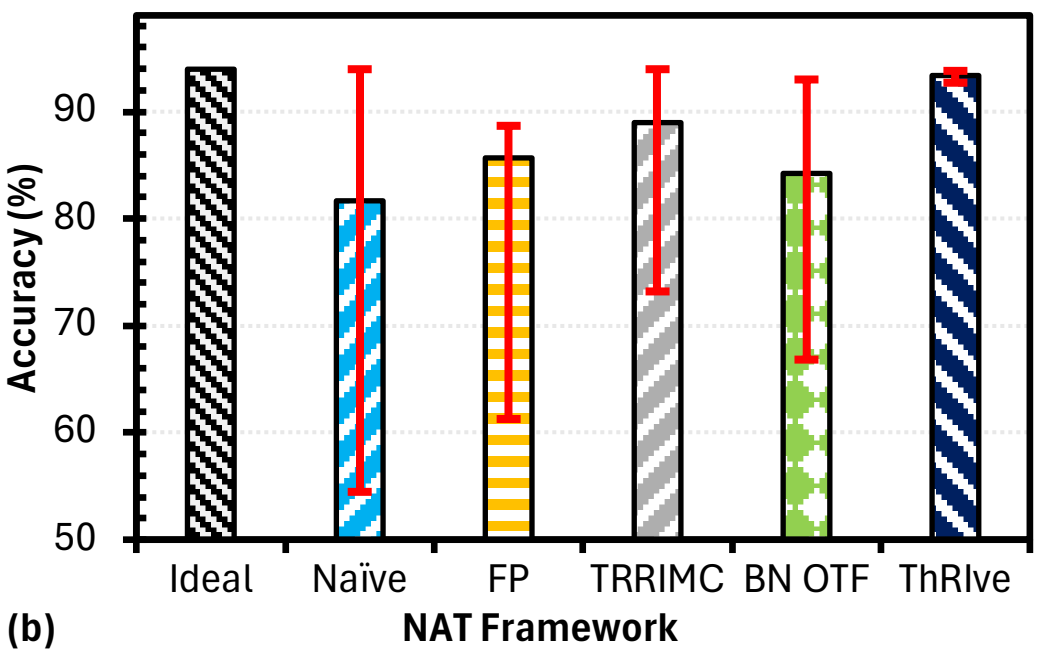

Figure 11: Generalized accuracy comparison on the CIFAR-10 dataset using ResNet18. (a) ThRIve's BO-based NAT framework on HyperX. (b) Comparison with existing NAT methods shows that ThRIve achieves consistent generalized accuracy (lower range deviation and high mean accuracy).

To address these issues, we incorporate a BO-based NAT framework that adaptively selects thermal noise levels for training. Unlike conventional approaches that rely on uniformly spaced or sequential noise levels, BO efficiently explores the entire thermal noise sample space and identifies the best noise levels to train on based on the min-max optimization formulation as discussed in Section 4.2. This addresses both key limitations of conventional training approaches: it avoids the need to predefine a large set of training points across a continuous temperature range, and it eliminates the reliance on a fixed training order. Fig. 11(a) shows the generalized accuracy achieved using ThRIve's BO-based NAT framework on HyperX (HyperX-T). As observed, the accuracy remains consistent across the entire operating temperature range. The range deviation stays below 2% of the mean accuracy over the full temperature range, which is also used as the convergence criterion in the ThRIve NAT framework. Fig. 11(b) compares ThRIve with existing NAT methodologies discussed in Section 5.2. The baselines exhibit high range deviation and lower mean accuracy, whereas ThRIve outperforms them by a significant margin. Note that both baselines and ThRIve are evaluated on HyperX. Fig. 11(b) also shows that heterogeneity alone isn't sufficient to provide thermal robustness as can be seen from the generalized accuracy achieved in the case of baseline NAT techniques.

### 5.5 ThRIve Performance Analysis across Hardware Architectures

This subsection demonstrates how ThRIve can be applied across different hardware architectures. We first evaluate its performance in terms of generalized accuracy and EDP on existing heterogeneous PIM setups. These include architectures such as HyperX, HyDE, and HuNT [12], [13], [19]. We consider three CNN models: VGG16, ResNet18, and MobileNetV3_Small, with the CIFAR-100 dataset for this analysis. Next, we show ThRIve's applicability across three diverse datasets: CIFAR-10, CIFAR-100, and TinyImageNet. To further enhance noise resilience, we incorporate BN, which helps reduce shifts in activation distributions. Following prior work, BN parameters (scale and shift) are trained offline, while the BN statistics (mean and variance) are updated online for each batch of data [10], [20]. This allows the model to better handle activation shifts caused by thermal noise during inference. Additionally, we consider an alternative configuration for each architecture, where all types of PIM devices in the system are replaced with SRAM PIM to create an all-SRAM variant (e.g., HyperX-S). This configuration serves as the ideal, noise-free reference, representing thermal noise-resilient performance. It also allows us to highlight the benefits of incorporating ThRIve into the original architecture. To ensure a fair comparison, all configurations for each individual architecture are constrained to the same chip area. However, as shown in Table I, SRAM cells occupy significantly higher area compared to NVM-based cells, resulting in substantially lower on-chip storage capacity for the all-SRAM variants. This storage capacity reduction leads to frequent DRAM access to fetch both the base model and LoRA parameters in scenarios when the weights cannot be stored entirely on-chip. The associated DRAM access latency and energy overhead to fetch these parameters are incorporated into our performance evaluations, ensuring our performance and energy evaluations accurately reflect the end-to-end system performance and energy during inference.

In our analysis, the original heterogeneous architecture is referred to as the naïve variant (e.g., HyperX-N). The version integrated with ThRIve is referred to as the ThRIve variant (e.g., HyperX-T). Note that there are no architectural differences between the naïve and ThRIve variants. The distinction lies solely in the NAT methodology employed. All architectures include a noise-resilient component, specifically, the SRAM-based PIM. ThRIve takes advantage of this by mapping the LoRA matrices onto the SRAM PIM (see Fig. 5), allowing for effective mitigation of thermal noise.

Fig. 12(a) presents the generalized accuracy of the HyperX architecture across all three variants: HyperX-N, HyperX-S, and HyperX-T. As observed, HyperX-N performs better than the sole ReRAM-based PIM architecture (see Fig. 3). This is because some layers in HyperX are mapped onto SRAM and are unaffected by thermal noise. However, layers mapped onto ReRAM do suffer from thermal noise. This leads to significant range deviation, as indicated by the error bars, and results in lower mean value of the accuracy across the entire operating temperature range. Although HyperX supports on-device retraining of weights stored in the SRAM block, the weights in ReRAM still vary with temperature. Addressing these variations would require repeated on-device retraining. Since temperature is a continuous variable, this quickly becomes impractical due to increased inference latency. In contrast, HyperX-T, achieves generalized accuracy nearly equal to the noise-free baseline, HyperX-S. This highlights the effectiveness of ThRIve as a NAT approach. Fig. 12(b) shows the normalized EDP values for all three configurations, with HyperX-S used as the reference. We employ EDP as the performance evaluation metric since it captures both energy and latency in a single term. Notably, HyperX-T delivers consistent accuracy across the entire thermal range. At the same time, it achieves up to a ~4.8× reduction in EDP compared to HyperX-S.

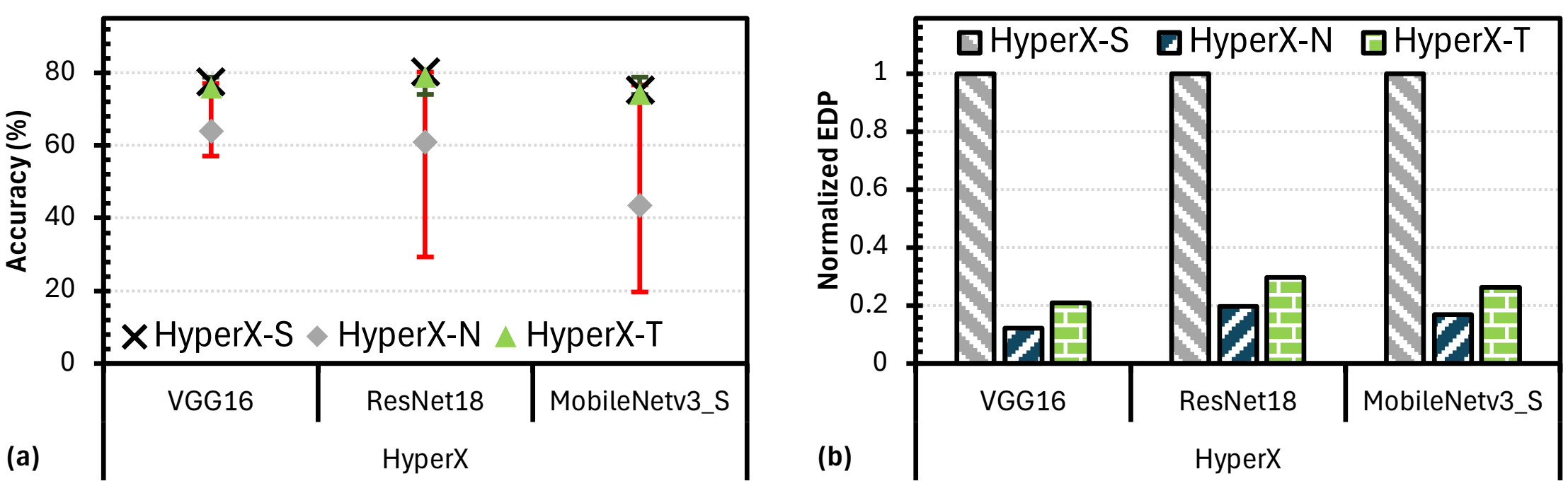

Figure 12: Comparison of (a) generalized accuracy and (b) normalized energy-delay product (EDP) for HyperX architecture across the three configurations: Naïve (HyperX-N), ThRIve-integrated (HyperX-T), and SRAM-only (HyperX-S).

Additionally, we assess ThRIve's robustness against spatial temperature variations by modeling the thermal profile of the HyperX architecture, capturing both horizontal and vertical heat flow, as shown in existing literature [58], [59]. HyperX was chosen as a representative heterogeneous PIM architecture, and ResNet18 with the CIFAR-100 dataset was used as the workload for this evaluation. We then injected thermal noise corresponding to this spatial temperature profile into the ThRIve-trained model (HyperX-T). The model achieved an accuracy (79.52%) within the 2% deviation band across the full operating temperature range, demonstrating that our methodology remains effective even under spatial thermal variations.

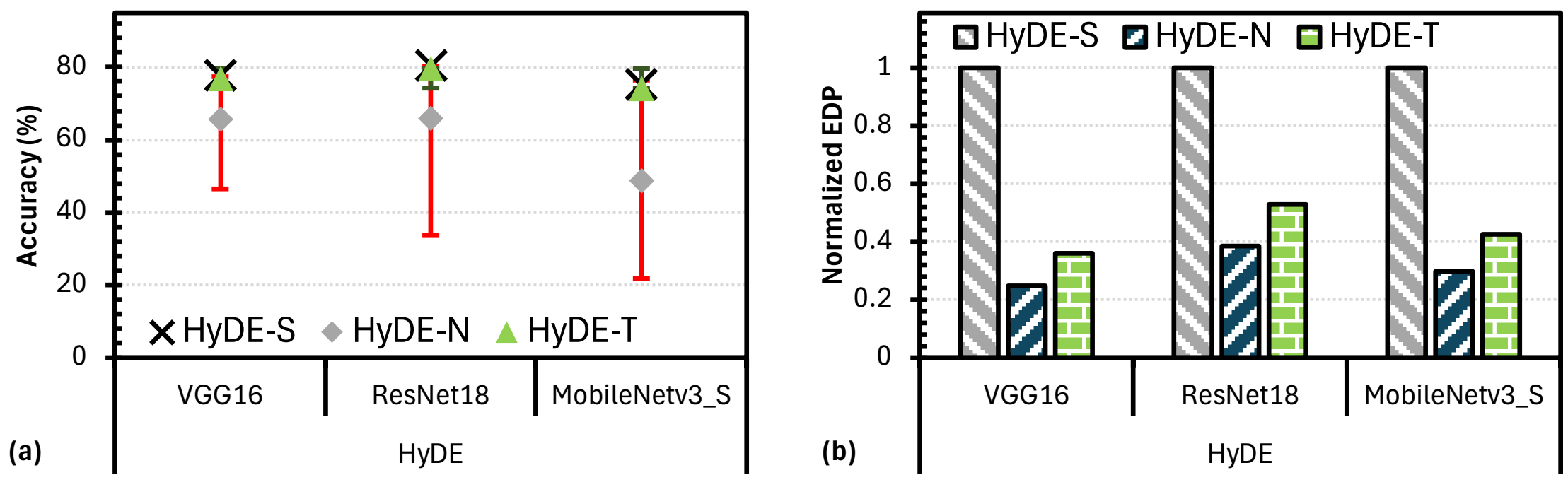

Figure 13: Comparison of (a) generalized accuracy and (b) normalized energy-delay product (EDP) for HyDE architecture across the three configurations: Naïve (HyDE-N), ThRIve-integrated (HyDE-T), and SRAM-only (HyDE-S).

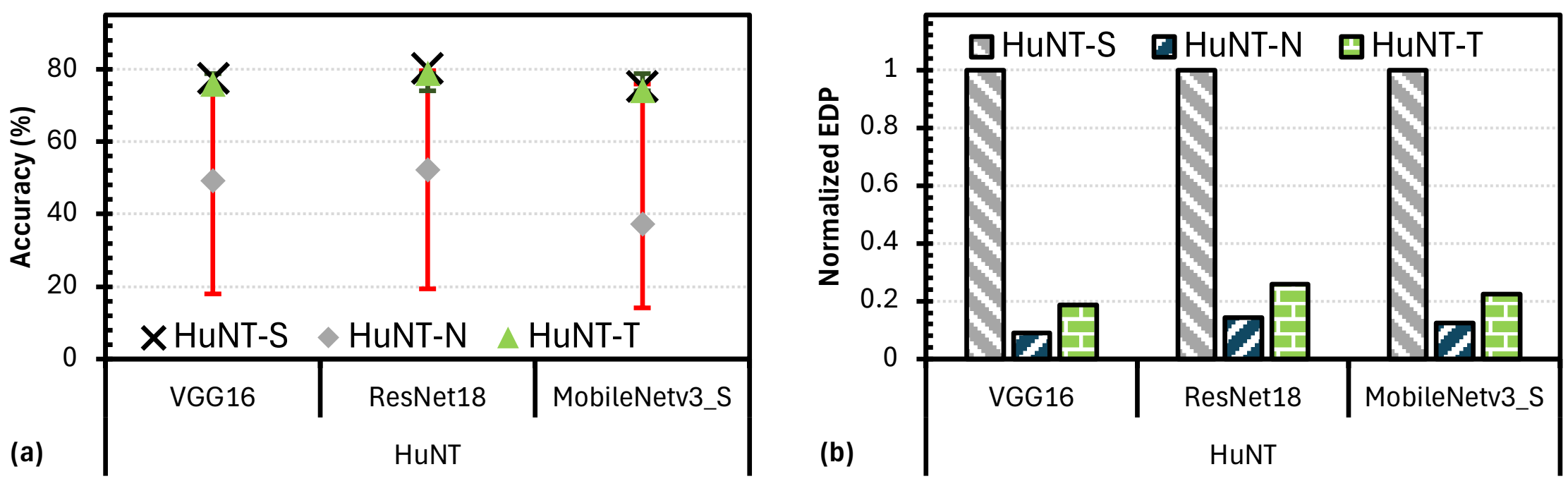

Figure 14: Comparison of (a) generalized accuracy and (b) normalized energy-delay product (EDP) for HuNT architecture across the three configurations: Naïve (HuNT-N), ThRIve-integrated (HuNT-T), and SRAM-only (HuNT-S).

Table 3: GPU runtime using block level details (epochs/iterations/time) for different models using CIFAR-100 dataset on HyDE.

| **Algorithm Variable** | **MobileNetv3_Small** | **ResNet18** | **VGG16** |
|---|---|---|---|
| Parameters (in Millions) | 1.62 | 11.17 | 15.13 |
| $A$ | 5 | 7 | 8 |
| $B$ | 34 | 37 | 44 |
| $t_{epoch}$ | 9.2 | 10.1 | 11.5 |
| $t_{initial}$ (in Seconds) | 0.92 | 1.7 | 2.4 |
| $t_{GP_OPS}$ (in Seconds) | 0. 88 | 0.97 | 1.26 |
| Total time (in Minutes) | ≈ 28 | ≈ 45 | ≈ 69 |

Next, we incorporate ThRIve into HyDE, an architecture that combines PCM and FeFET-based PIM with SRAM-based PIM. HyDE selects the optimal device type out of PCM, FeFET, or SRAM for each DNN layer to balance accuracy, area, and energy under device-level non-idealities. We also integrate ThRIve into HuNT, which uses ReRAM, FeFET, and SRAM-based PIM. HuNT enables energy-efficient training by computing gradients on the SRAM tier. However, the architecture and training process do not account for the impact of thermal noise. Fig. 13(a, b) and Fig. 14(a, b) presents the generalized accuracy and EDP of HyDE and HuNT across all three configurations. As can be observed from Fig. 13, HyDE-T achieves accuracy at par with the noise-free HyDE-S baseline while reducing EDP by up to ~2.8×. From the results in Fig. 14, we observe that HuNT-T delivers consistent accuracy nearly equal to HuNT-S, with up to ~5.4× savings in EDP.

From Fig. 12, 13, and 14, we can also observe that the ThRIve variants incur a notable overhead compared to the naïve counterparts. This overhead is necessary for achieving thermal robustness. Without ThRIve, the architectures remain prone to thermal noise and fail to maintain consistent accuracy when exposed to it. Thermal noise is a critical challenge in such systems. If not properly addressed, it can make these designs impractical and unusable in real-world scenarios. The overhead introduced by ThRIve is the necessary cost to ensure consistent inference accuracy across the entire operating temperature range. In addition, we report the runtime of the ThRIve algorithm for MobileNetv3_small, ResNet18, and VGG16 on the CIFAR-100 dataset using Equation 14 in Table 3. Here, the values of $A$ (epochs consumed per training session), $B$ (BO iterations to convergence), and $t_{epoch}$ (wall-clock time per epoch) depends on the workload (model & dataset) characteristics. Unlike conventional noise-aware training approaches that require retraining entire model parameters, our framework leverages LoRA's parameter-efficient fine-tuning paradigm, where only low-rank decomposition matrices are trainable while the original pre-trained weights remain frozen. This ensures that our approach scales effectively with the model size where ResNet18 and VGG16 have about 7× and 10× more parameters than MobileNetv3_Small (see Table 3).

Next, we show ThRIve's ability to provide consistent accuracy across different datasets. For this analysis, we have considered the HyDE architecture as an example. Fig. 15 shows the mean and range deviation of the generalized accuracy achieved for different DNN models across various datasets including CIFAR-10 (C-10), CIFAR-100 (C-100), and TinyImageNet (TINet). From the figure, we can observe that the HyDE-N variant shows a noticeable degradation in accuracy. This results in a larger range deviation and a lower mean accuracy. In contrast, the HyDE-T variant achieves accuracy with a very small range deviation and a mean value that is very close to the SRAM-only variant, HyDE-S. Overall, these evaluations using various architectures, DNN models, and datasets demonstrate the broad applicability of ThRIve in enabling consistent and thermally robust inference on PIM based systems.

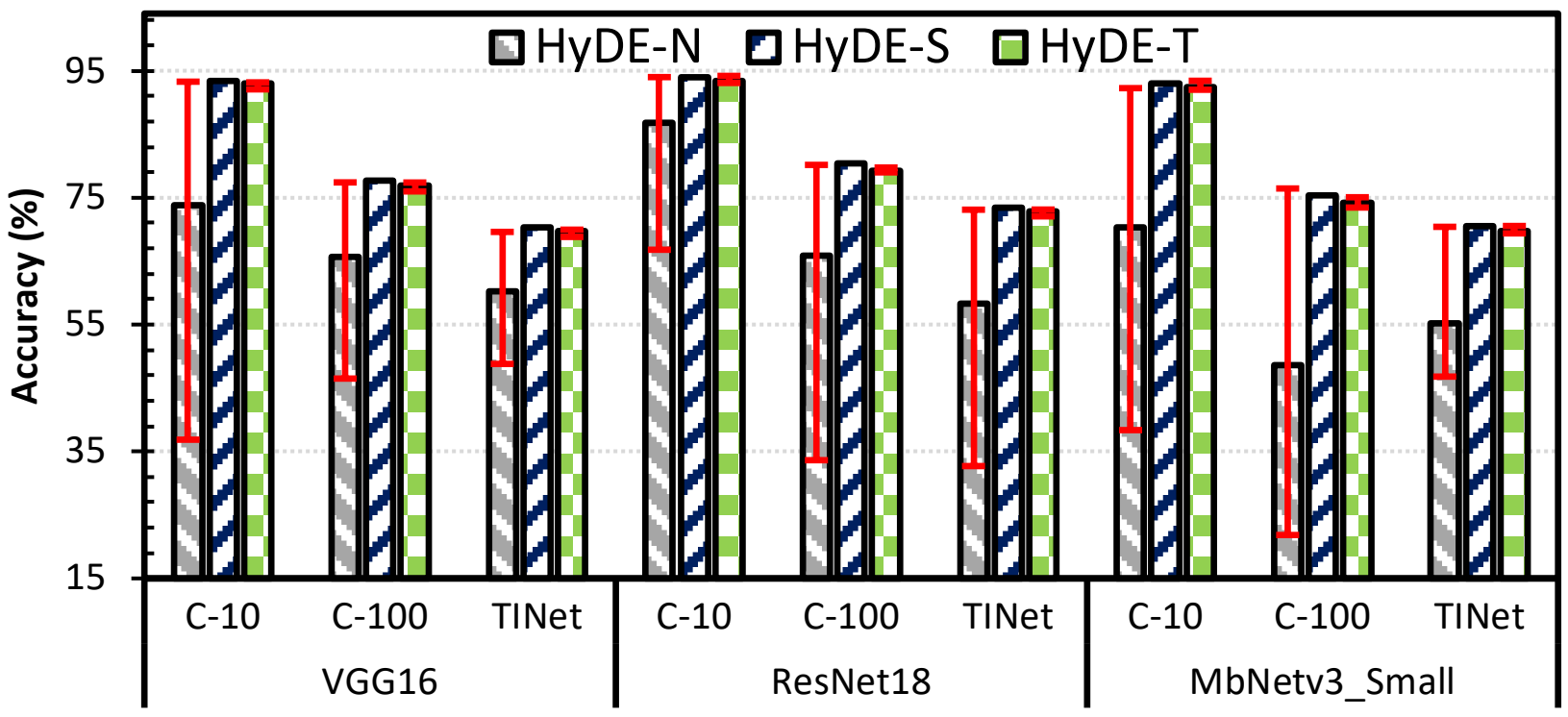

Figure 15: Comparison of generalized accuracy across three configurations: Naïve (HyDE-N), ThRIve-integrated (HyDE-T), and SRAM-only (HyDE-S), for multiple DNN models evaluated on CIFAR-10 (C-10), CIFAR-100 (C-100), and TinyImageNet (TINet) datasets. Results show both mean of generalized accuracy and range deviation across the full operating temperature range.

## 6 CONCLUSION

PIM architectures featuring non-volatile memories such as ReRAM, PCM, FeFET, etc., have gained significant traction for accelerating ML workloads at the edge. Despite their efficiency in performing in-memory computation, these architectures are highly susceptible to thermal noise, which distorts the saved model parameters and leads to a drop in model inference accuracy. In this work, we introduced ThRIve, a noise-aware training framework that leverages low rank adaptation with heterogeneous PIM architectures to enable robust inference under temperature-induced noise. ThRIve leverages a hardware and software co-design strategy, selectively storing trainable low-rank parameters on thermally resilient SRAM while the model's original weights remain on energy-efficient NVM such as ReRAM. By doing so, ThRIve efficiently mitigates thermal noise without requiring frequent on-device retraining. Experimental evaluations across multiple architectures, models, and datasets demonstrate that ThRIve consistently delivers high inference accuracy across the entire operating temperature range. It maintains a mean accuracy within 2% of the noise-free baseline and keeps the range deviation under 2% of the mean. Furthermore, ThRIve achieves significant energy-delay product (EDP) savings, up to ~4.8× on HyperX, ~2.8× on HyDE, and ~5.4× on HuNT, compared to their respective noise-free SRAM-only baselines during model inferencing, offering a thermally robust and energy-efficient alternative for implementing ML on edge using PIM based systems.